\documentclass[manuscript]{aa}
\usepackage{graphicx,booktabs,multirow}
 \usepackage{amsmath}

\def\o{$^{\rm o}$}

\authorrunning{Wang et al.}

\begin{document}
\title{GLOBULAR CLUSTERS IN THE OUTER HALO OF M31\thanks{Tables 1 and 2 are only available in electronic form at the CDS via anonymous ftp to cdsarc.u-strasbg.fr (130.79.128.5) or via http://cdsweb.u-strasbg.fr/cgi-bin/qcat?J/A+A/}}
\titlerunning{Halo clusters in M31}

\author{
Song Wang\inst{1}, Jun Ma\inst{1,2}, AND Jifeng Liu\inst{1,2,3}}

\institute{
$^1$ Key Laboratory of Optical Astronomy, National Astronomical Observatories, Chinese
Academy of Sciences, Beijing 100101, China; songw@bao.ac.cn\\
$^2$ College of Astronomy and Space Sciences, University of Chinese Academy of Sciences, Beijing 100049, China\\
$^3$ WHU-NAOC Joint Center for Astronomy, Wuhan University, Wuhan, Hubei 430072, China
}

\abstract
{In this paper, we present photometry of 53 globular clusters (GCs) in the M31 outer halo, including the {\sl GALEX} FUV and NUV, SDSS $ugriz$, 15 intermediate-band filters of BATC, and 2MASS $JHK_{\rm s}$ bands.
By comparing the multicolour photometry with stellar population synthesis models, we determine the metallicities, ages, and masses for these GCs, aiming to probe the merging/accretion history of M31.
We find no clear trend of metallicity and mass with the de-projected radius. The halo GCs with age younger than $\approx$ 8 Gyr are mostly located at the de-projected radii around 100 kpc, but this may be due to a selection effect.
We also find that the halo GCs have consistent metallicities with their spatially-associated substructures, which provides further evidence of the physical association between them.
Both the disk and halo GCs in M31 show a bimodal luminosity distribution.
However, we should emphasize that there are more faint halo GCs which are not being seen in the disk.
The bimodal luminosity function of the halo GCs may reflect different origin or evolution environment in their original hosts.
The M31 halo GCs includes one intermediate metallicity group ($-1.5 <$ [Fe/H] $< -0.4$) and one metal-poor group ([Fe/H] $<-1.5$), while the disk GCs have one metal-rich group more.
There are considerable differences between the halo GCs in M31 and the Milky Way (MW).
The total number of M31 GCs is approximately three times more numerous than that of the MW, however,
M31 has about six times the number of halo GCs in the MW.
Compared to M31 halo GCs, the Galactic halo ones are mostly metal-poor.
Both the numerous halo GCs and the higher-metallicity component are suggestive of an active merger history of M31.}

\keywords{galaxies: individual (M31) -- galaxies: globular clusters --
galaxies: stellar content}

\maketitle

\section{INTRODUCTION}

The formation and evolution history of galaxies are not quiescent.
In hierarchical cosmological models \citep{White1978}, galaxies grow in mass through continual interaction with nearby smaller satellites.
Due to the process of phase mixing and the faint feature of the debris, however, it is not simple for us to probe these merging/accretion events occurred at distant past.
Generally, there are two ways to study the interaction history of  galaxies:
(1) modeling galaxy interactions to reproduce the substructures discovered in the outskirts of the galaxy;
(2) studying the kinematics and stellar populations (e.g., metallicity, age) of the halo components.

Globular clusters (GCs) in the halo of galaxies are often considered as debris of the galaxies' interaction history.
Thus, they are valuable targets for studying this topic.
In the Milky Way (MW), it has long been noticed that some GCs were depositing into the halo \citep[e.g.,][]{Ibata1994,DaCosta1995,Martin2004} from disrupting dwarfs (e.g., Sagittarius, Canis Major),
although the Canis Major overdensity may also be produced by the thin and thick disk and spiral arm populations of the MW rather than by a collision with a dwarf satellite galaxy \citep[e.g.,][]{Mateu2009}.
Some of the surviving dwarf galaxies associated with the MW lie in a rotating planar structure \citep{Metz2007}, and many GCs in the Galactic halo share this planar alignment \citep{Keller2012, Pawlowski2012}.
These directly confirm the basic theory of the hierarchical galaxy formation model \citep{McConnachie2009}, and it is important to ascertain whether this assembly mode is prevalent in other galaxies.

An attractive alternative is our nearest large neighbor, the Andromeda Galaxy (M31), which includes more GCs than the MW system (by a factor of $\approx$ 3). Some surveys have discovered many substructures in the halo of M31 \citep[see][for details]{Ferguson2002,Ibata2007,McConnachie2009,McConnachie2018}. Recently, the Pan-Andromeda Archaeological Survey (PAndAS) confirmed numerous halo GCs \citep[e.g.,][]{Huxor2014}, and the halo GCs show a flat radial surface-density profile beyond the projected distance $R_{\rm p} \approx 30$ kpc \citep{Huxor2011}.
Many of them are spatially associated with halo substructures.
Using Monte Carlo simulations, \citet{Mackey2010} argued a genuine physical association between the halo GCs and multiple tidal debris streams.
Detailed kinematics analysis also shows that the GCs with $R_{\rm p} > 30$ kpc exhibit coherent rotation around the minor optical axis of M31 \citep{Veljanoski2014},
which appears to be rotating in the same direction as some dwarf galaxies associated with M31 \citep{Ibata2013}.
All of these discoveries have been interpreted as evidence of an accretion process \citep{Cote2000,Huxor2011}.
However, a major limitation of previous studies is the lack of comparison of stellar population (e.g., age, metallicity) between halo GCs and these substructures, which may provide strong evidence of the physical association.

In this paper, we provide photometry of a set of GCs in the M31 outer halo, using images obtained with the Beijing-Arizona-Taiwan-Connecticut (BATC) Multicolour Sky Survey Telescope.
We construct observed spectral energy distributions (SEDs) from 1538 to 20,000 \AA by combining the {\sl GALEX} (FUV and NUV), SDSS $ugriz$, BATC, and Two Micron All Sky Survey (2MASS) NIR data.
Comparing the SEDs with simple stellar population (SSP) models, we estimate the metallicities, ages, and masses for the GCs. This paper is organized as follows. In Section 2, we present the BATC observations of the sample clusters and steps of relevant data-processing and photometry. In Section 3, we derive the metallicities, ages, and masses of the sample clusters. In Section 4, we discuss the properties of the halo GCs by studying the association between them and halo substructures, and by comparing them with the M31 disk GCs and Galactic halo GCs.
Finally, we summarize our results in Section 5.

\section{SAMPLE, OBSERVATIONS AND DATA REDUCTION}

\subsection{Sample and BATC Observations}

\begin{figure}
\center
\includegraphics[width=0.47\textwidth]{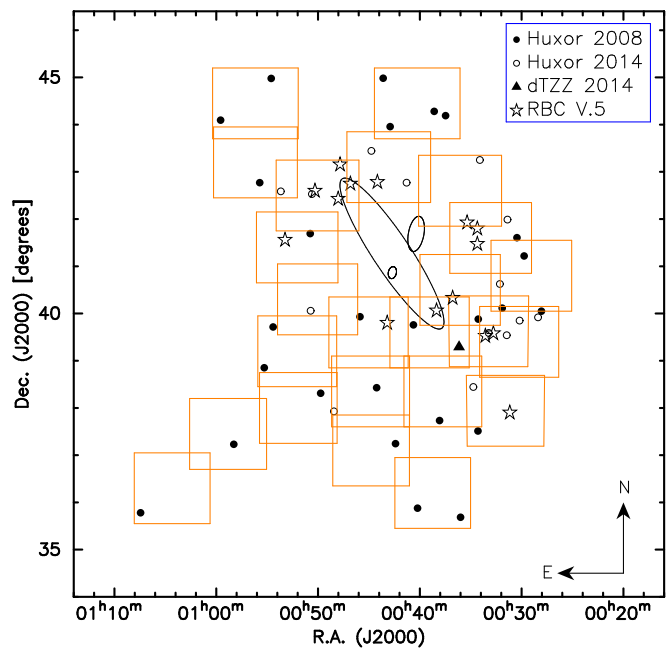}
\caption{Spatial distribution of 56 observed GCs. A box represents a field view of 1.5\o $\times $ 1.5\o. The large ellipse has a semimajor axis of 2\o (27 kpc), representing a disk with an inclination of 77.5\o and a position angle of 38\o \citep{ken89}. The M31 optical disk lies well within this ellipse. The two small ellipses are the $D_{25}$ isophotes of NGC 205 (northwest) and M32 (southeast).}
\label{spatial.fig}
\end{figure}

We selected GCs in the outer halo of M31,
with projected distance $R_{\rm p} >$ 20 kpc,
from \citet{Huxor2008,Huxor2014}, \cite{dTZZ2013,dTZZ2014}, and the RBC V.5 catalogue \citep{Galleti2004}.
The nonclusters were excluded with new classifications \citep{Caldwell2009,Huxor2014}.
Although the cluster G339 was classified as ``STAR'' in \citet{Caldwell2009}, the {\it HST} observation showed that it is a true GC \citep{Barmby2007};
we included it in our sample.
This led to 113 outer halo GCs, of which 56 objects are located in the field of BATC observations (Figure \ref{spatial.fig}).

The BATC 60/90 cm Schmidt Telescope is at the Xinglong Station of the National Astronomical Observatories, Chinese Academy of Sciences (NAOC).
The telescope includes 15 intermediate-band filters,
covering a range of wavelength from 3000 to 10,000 \AA~\citep[see][for details]{Fan1996}.
The CCD size is $4{\rm k}\times4{\rm k}$, with a resolution of $1''.36$ pixel$^{-1}$ \citep{Fan2009}.
%

\subsection{BATC Photometry}

\begin{figure}
\center
\includegraphics[width=0.47\textwidth]{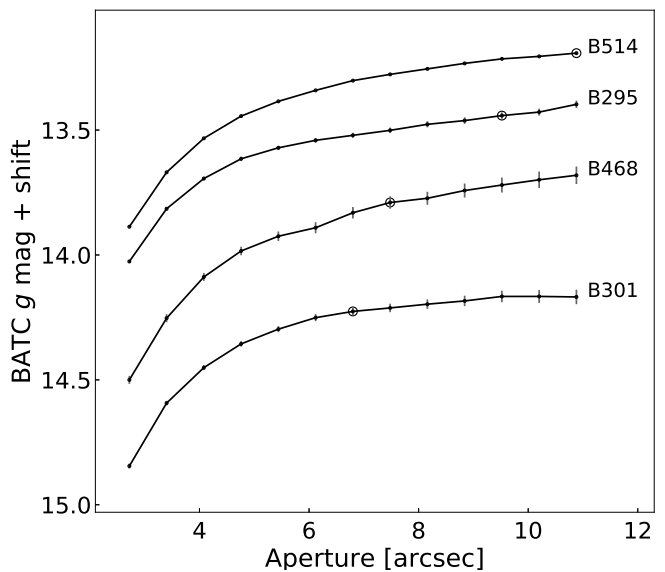}
\caption{Curve of growth using the BATC $g$-band photometry for several GCs as an example. The open circles represent the radii adopted for aperture photometry.}
\label{growth.fig}
\end{figure}

\begin{figure*}
\center
\includegraphics[width=0.98\textwidth]{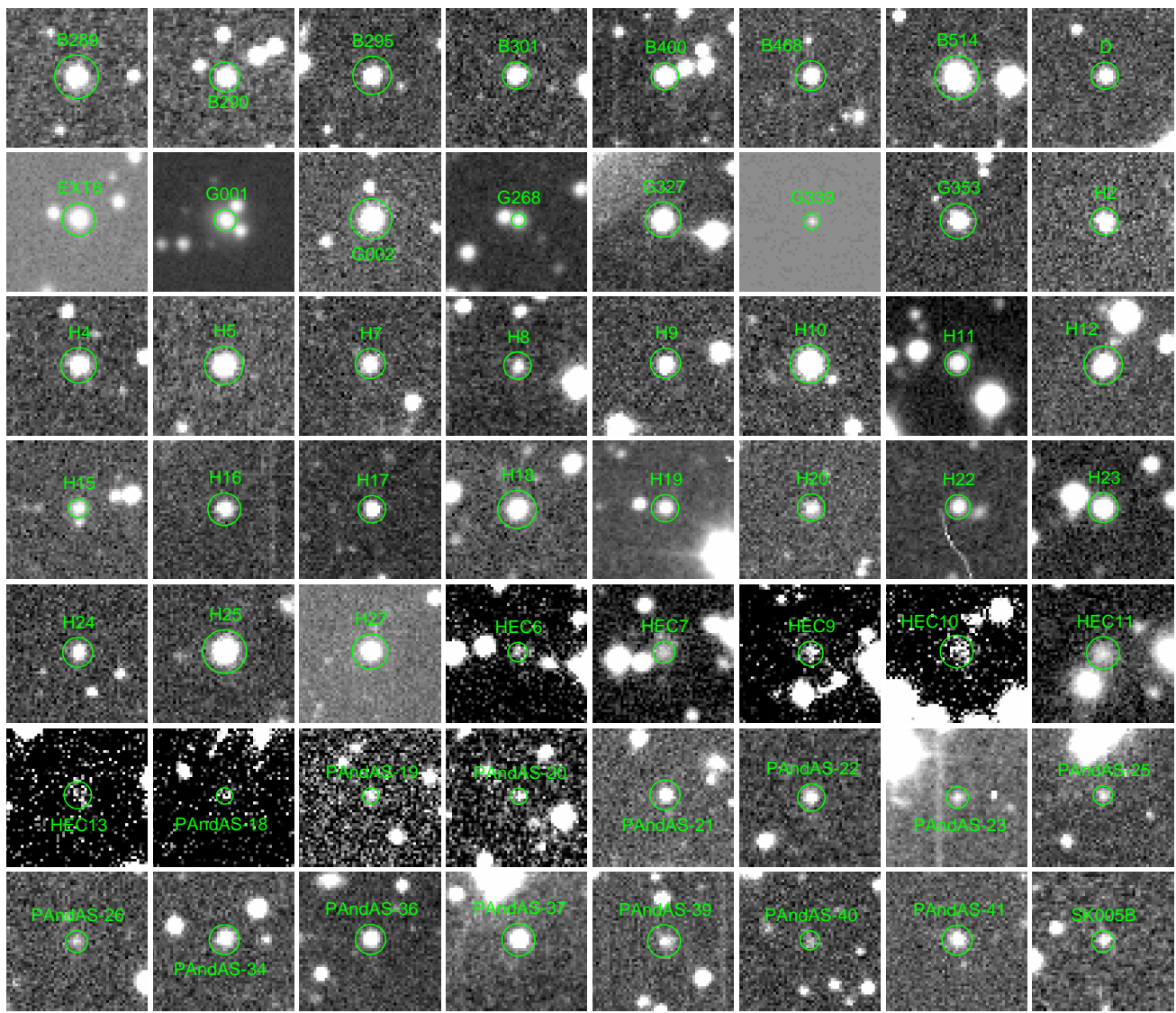}
\caption{ Finding charts of 56 GCs in the
BATC $g$ band, obtained with the NAOC 60/90cm Schmidt telescope. The field of view of each image is $\approx$ $2^{\prime}\times2^{\prime}$.
The green circles represent the radii for aperture photometry.}
\label{ds9.fig}
\end{figure*}

We processed all the CCD images with standard procedures including bias subtraction and flat fielding \citep{Fan1996,Zheng1999}.
In order to improve the image quality, we combined multiple images of the same filter to one and performed photometry using the combined images.
The BATC magnitudes were defined and obtained in a similar way as for the spectrophotometric AB magnitude system \citep[e.g.,][]{Ma2009}.
We performed standard aperture photometry of our sample clusters using the {\sc PHOT} routine in {\sc DAOPHOT} \citep{Stetson1987}.
To ensure that we adopted the most appropriate photometric radius ($R_{\rm ap}$) that includes all light from the object,
we produced a curve of growth from the $g$-band photometry obtained through apertures with radii in the
range of 2 -- 8 pixel ($\approx$ 2.7 -- 10.9 arcsec) with 0.5 pixel increments
(Figure \ref{growth.fig}).
In order not to include the light from extraneous objects, we also checked the aperture radii carefully by visual examination of the images.
This method ensures that we can correctly determine the total luminosities of clusters,
especially as the use of small apertures for small clusters can maximize the signal-to-noise ratio (S/N) and minimize the contamination from nearby sources \citep{Ma2015}.
The local sky background was measured in an annulus with an inner radius of $R_{\rm ap}+2$ pixel and a width of 5 pixels.
Finding charts of the GCs and the final photometric radii are shown in Figure \ref{ds9.fig}.

Three clusters (HEC6, PAndAS-18, and PAndAS-40) are too faint that we cannot derive accurate photometry, and we removed them from the sample.
The BATC photometry for the 53 GCs are listed in Table \ref{batc.tab}. For some objects, the magnitudes in some filters could not be obtained due to the low S/N. The magnitudes with an uncertainty larger than 0.3 would not be used in
following SED fitting, although they are listed in Table \ref{batc.tab}.

\begin{table*}[htp!]
\begin{center}
\small
\setlength{\tabcolsep}{0.3em}
\renewcommand\arraystretch{1.15}
\caption{BATC intermediate-band photometry of 53 sample GCs in M31.} \vspace{0mm}
\label{batc.tab}
\begin{tabular}{lcccccccccccccccc}
\hline\noalign{\smallskip}
Object & $a$ & $b$ & $c$ & $d$ & $e$ & $f$ & $g$ & $h$ & $i$ & $j$ &$ k$ & $m$ & $n$ & $o$ & $p$ & $R_{\rm ap}$\\
       &(mag)&(mag)&(mag)&(mag)&(mag)&(mag)&(mag)&(mag)&(mag)&(mag)&(mag)&(mag)&(mag)&(mag)&(mag)& ($''$)  \\
  (1) & (2) &(3) &(4) &(5) &   (6) & (7) &(8) &(9) &(10) &  (11) & (12) &(13) &(14) &(15) &  (16) & (17) \\
\hline\noalign{\smallskip}
B289      & 17.96     & 17.12     & 16.72     & 16.45     & 16.34     & 16.15     & 15.95     & 15.89     & 15.71     & 15.66     & 15.68     & 15.56     & 15.54     & 15.33     & 15.45     & 10.9\\
          & 0.06      & 0.04      & 0.02      & 0.04      & 0.02      & 0.01      & 0.02      & 0.01      & 0.03      & 0.01      & 0.02      & 0.02      & 0.02      & 0.02      & 0.05      &\\
B290      & 19.26     & 18.74     & 18.05     & 17.76     & 17.44     & 17.24     & 17.0      & 16.88     & 16.65     & 16.61     & 16.57     & 16.46     & 16.4      & 16.15     & 16.23     & 7.5\\
          & 0.12      & 0.06      & 0.03      & 0.05      & 0.02      & 0.02      & 0.03      & 0.02      & 0.03      & 0.02      & 0.02      & 0.03      & 0.02      & 0.03      & 0.07      &\\
B295      & 18.36     & 17.59     & 17.49     & 16.92     & 16.93     & 16.74     & 16.57     & 16.45     & 16.33     & 16.29     & 16.24     & 16.18     & 16.16     & 16.06     & 15.83     & 9.5\\
          & 0.07      & 0.04      & 0.03      & 0.03      & 0.02      & 0.02      & 0.01      & 0.01      & 0.01      & 0.01      & 0.02      & 0.03      & 0.02      & 0.02      & 0.07      &\\
B301      & 19.4      & 18.8      & 18.26     & 17.55     & 17.44     & 17.27     & 16.99     & 16.85     & 16.66     & 16.6      & 16.42     & 16.39     & 16.27     & 16.24     & 16.15     & 6.8\\
          & 0.13      & 0.05      & 0.04      & 0.03      & 0.02      & 0.01      & 0.02      & 0.01      & 0.03      & 0.02      & 0.02      & 0.03      & 0.03      & 0.04      & 0.18      &\\
B400      & 19.03     & 17.89     & 17.43     & 17.13     & 16.78     & 16.57     & 16.27     & 16.21     & 16.08     & 16.04     & 15.81     & 15.79     & 15.75     & 15.78     & 15.79     & 6.8\\
          & 0.08      & 0.04      & 0.03      & 0.01      & 0.02      & 0.02      & 0.03      & 0.02      & 0.02      & 0.02      & 0.02      & 0.01      & 0.02      & 0.02      & 0.04      &\\
B468      & ...       & 19.23     & 18.87     & 18.42     & 18.34     & 18.12     & 17.78     & 17.63     & 17.52     & 17.44     & 17.03     & 17.28     & 17.16     & 17.32     & 16.58     & 7.5\\
          & ...       & 0.07      & 0.05      & 0.04      & 0.02      & 0.03      & 0.03      & 0.03      & 0.04      & 0.04      & 0.04      & 0.04      & 0.06      & 0.06      & 0.17      &\\
B514      & 17.77     & 16.71     & 16.54     & 16.28     & 16.09     & 15.9      & 15.68     & 15.59     & 15.48     & 15.36     & 15.28     & 15.23     & 15.17     & 15.15     & 14.74     & 10.9\\
          & 0.07      & 0.06      & 0.06      & 0.01      & 0.02      & 0.03      & 0.03      & 0.01      & 0.02      & 0.01      & 0.03      & 0.03      & 0.04      & 0.03      & 0.05      &\\
SDSS-D         & 19.59     & 18.93     & 18.5      & 17.94     & 17.85     & 17.71     & 17.47     & 17.37     & 17.22     & 17.16     & 17.05     & 16.96     & 16.93     & 17.1      & 16.5      & 6.8\\
          & 0.16      & 0.05      & 0.04      & 0.03      & 0.02      & 0.01      & 0.03      & 0.02      & 0.03      & 0.02      & 0.03      & 0.04      & 0.04      & 0.06      & 0.15      &\\
EXT8      & 17.01     & 16.23     & 16.31     & 15.97     & 15.89     & 15.68     & 15.5      & 15.38     & 15.29     & 15.33     & 15.0      & 15.02     & 15.08     & 15.0      & 15.13     & 8.2\\
          & 0.05      & 0.03      & 0.02      & 0.04      & 0.04      & 0.02      & 0.02      & 0.02      & 0.02      & 0.02      & 0.01      & 0.01      & 0.01      & 0.04      & 0.03      &\\
G001      & 15.98     & 15.38     & 15.01     & 14.19     & 14.18     & 14.01     & 13.71     & 13.64     & 13.42     & 13.36     & 13.23     & 13.37     & 13.13     & 13.18     & 13.04     & 5.4\\
          & 0.01      & 0.05      & 0.03      & 0.01      & 0.01      & 0.01      & 0.01      & 0.01      & 0.02      & 0.03      & 0.01      & 0.07      & 0.02      & 0.01      & 0.02      &\\
\hline
\end{tabular}
\end{center}
{NOTE: Column 1 gives the GC names. Columns 2 -- 16 present the magnitudes in the 15 BATC passbands. The $1\sigma$ magnitude uncertainties from {\sc DAOPHOT} are listed for each object on the second line
for corresponding passbands. Column 17 is the photometric apertures adopted in this paper.}\\
{(This table is available in its entirety in machine-readable form.)}
\end{table*}

\begin{table*}[htp!]
\begin{center}
\small
\renewcommand\arraystretch{1.15}
\caption{{\sl GALEX}, SDSS,
and 2MASS NIR photometry of 53 sample GCs in M31.} \vspace{3mm} \label{other.tab}
\begin{tabular}{lcccccccccc}
\hline\noalign{\smallskip}
Object& FUV & NUV & $u$ & $g$ & $r$ & $i$ & $z$ & $J$ & $H$ & $K_{\rm s}$ \\
      &(mag)&(mag)&(mag)&(mag)&(mag)&(mag)&(mag)&(mag)&(mag)&(mag)\\
       (1) & (2) &(3) &(4) &(5) &   (6) & (7) &(8) &(9) &(10)  & (11)\\
\hline\noalign{\smallskip}
B289      & 20.73     & 19.82     & ...       & ...       & ...       & ...       & ...       & 15.19     & 15.24     & 15.88     \\
          & 0.05      & 0.02      & ...       & ...       & ...       & ...       & ...       & 0.04      & 0.05      & 0.09      \\
B290      & 23.81     & 22.43     & ...       & ...       & ...       & ...       & ...       & 16.09     & 15.9      & 15.9      \\
          & 0.34      & 0.1       & ...       & ...       & ...       & ...       & ...       & 0.06      & 0.06      & 0.06      \\
B295      & 21.38     & 20.44     & 18.25     & 16.96     & 16.44     & 16.16     & 15.99     & 15.79     & 15.99     & 16.52     \\
          & 0.11      & 0.01      & 0.04      & 0.01      & 0.01      & 0.01      & 0.03      & 0.06      & 0.08      & 0.14      \\
B301      & 23.3      & 22.14     & 19.18     & 17.54     & 16.82     & 16.49     & 16.24     & 15.94     & 15.57     & 15.97     \\
          & 0.24      & 0.04      & 0.06      & 0.02      & 0.01      & 0.01      & 0.02      & 0.05      & 0.05      & 0.07      \\
B400      & 22.45     & 21.16     & 18.54     & 16.89     & 16.24     & 15.88     & 15.74     & 15.41     & 15.38     & 15.56     \\
          & 0.17      & 0.05      & 0.04      & 0.01      & 0.01      & 0.01      & 0.02      & 0.03      & 0.04      & 0.05      \\
B468      & 23.26     & 22.88     & 19.85     & 18.3      & 17.69     & 17.36     & 17.23     & 17.17     & 16.88     & 16.25     \\
          & 0.33      & 0.2       & 0.11      & 0.03      & 0.02      & 0.02      & 0.06      & 0.14      & 0.13      & 0.1       \\
B514      & 20.61     & 19.54     & 17.73     & 16.13     & 15.54     & 15.25     & 15.13     & 15.08     & 14.68     & 15.47     \\
          & 0.06      & 0.03      & 0.03      & 0.01      & 0.01      & 0.01      & 0.01      & 0.04      & 0.04      & 0.07      \\
SDSS-D         & ...       & 20.55     & 19.29     & 17.9      & 17.34     & 17.03     & 17.03     & 17.06     & 16.66     & 17.27     \\
          & ...       & 0.09      & 0.07      & 0.02      & 0.02      & 0.02      & 0.04      & 0.11      & 0.1       & 0.18      \\
EXT8      & ...       & 18.69     & 16.99     & 15.84     & 15.51     & 15.15     & 14.97     & 14.89     & 14.84     & 15.46     \\
          & ...       & 0.06      & 0.02      & 0.01      & 0.01      & 0.01      & 0.01      & 0.02      & 0.03      & 0.05      \\
G001      & 19.09     & 18.25     & 16.04     & 14.22     & ...       & 13.26     & 13.04     & 12.71     & 12.51     & 12.85     \\
          & 0.02      & 0.01      & 0.01      & 0.01      & ...       & 0.01      & 0.01      & 0.01      & 0.01      & 0.01      \\
\hline
\end{tabular}
\end{center}
{NOTE: Column 1 gives the GC names. Columns 2 -- 11 present the magnitudes from {\sl GALEX}, SDSS, and 2MASS NIR observations. All  magnitudes are in the AB magnitude system. The $1\sigma$ magnitude uncertainties from {\sc DAOPHOT} are listed for each object on the second line for the corresponding passbands.}\\
{(This table is available in its entirety in machine-readable form.)}
\end{table*}

\subsection{{\sl GALEX} UV, SDSS, and 2MASS NIR Photometry}

Fitting SEDs with SSP models is a general way to determine the ages and masses for star clusters. Accurate and numerous photometric data can help derive accurate ages \citep{deGrijs2003, Anders2004}. The UV photometry is powerful for age estimation of young stellar populations, and the combination of UV photometry with optical observations enables one to break the age--metallicity degeneracy \citep{Kaviraj2007,Fan2017}.
Also, the age--metallicity degeneracy can be partially broken by adding NIR photometry to optical colours \citep{deJong1996,Anders2004,Ma2009}.

We downloaded the $GALEX$, SDSS, and 2MASS images
and performed photometry using the same radii (in arcsec) following the BATC photometry.
For some clusters, accurate photometric measurements cannot be obtained for some bands due to various reasons.
For example, G001 is saturated in the SDSS $r$-band observation; H19 is located on the asterism of one nearby star in the SDSS $i$- and $z$-band observations.
To examine the quality and reliability of our photometry, we compared the aperture magnitudes of the 53 objects obtained here with previous results.

\begin{figure}
\center
\includegraphics[width=0.24\textwidth]{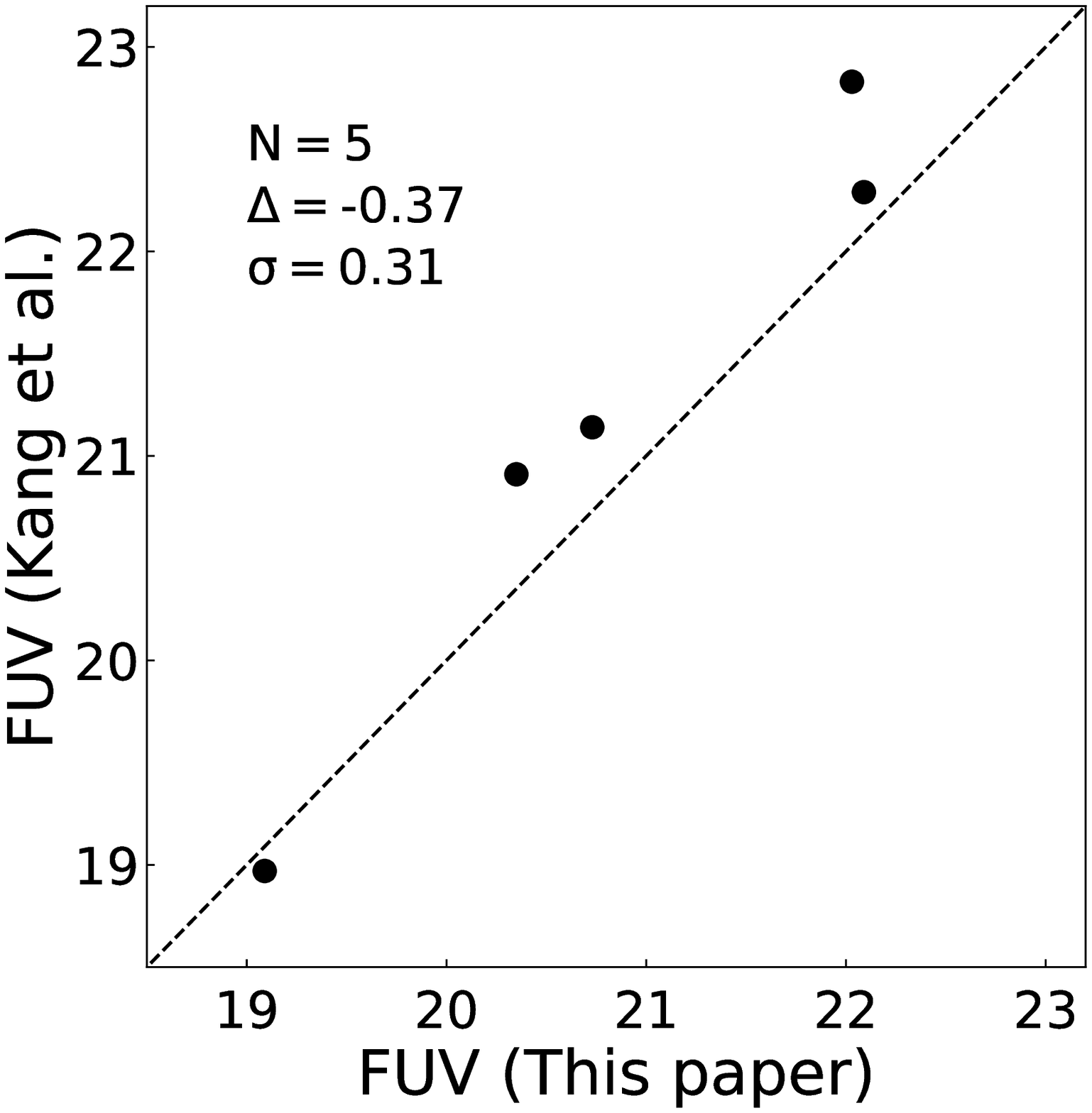}
\includegraphics[width=0.24\textwidth]{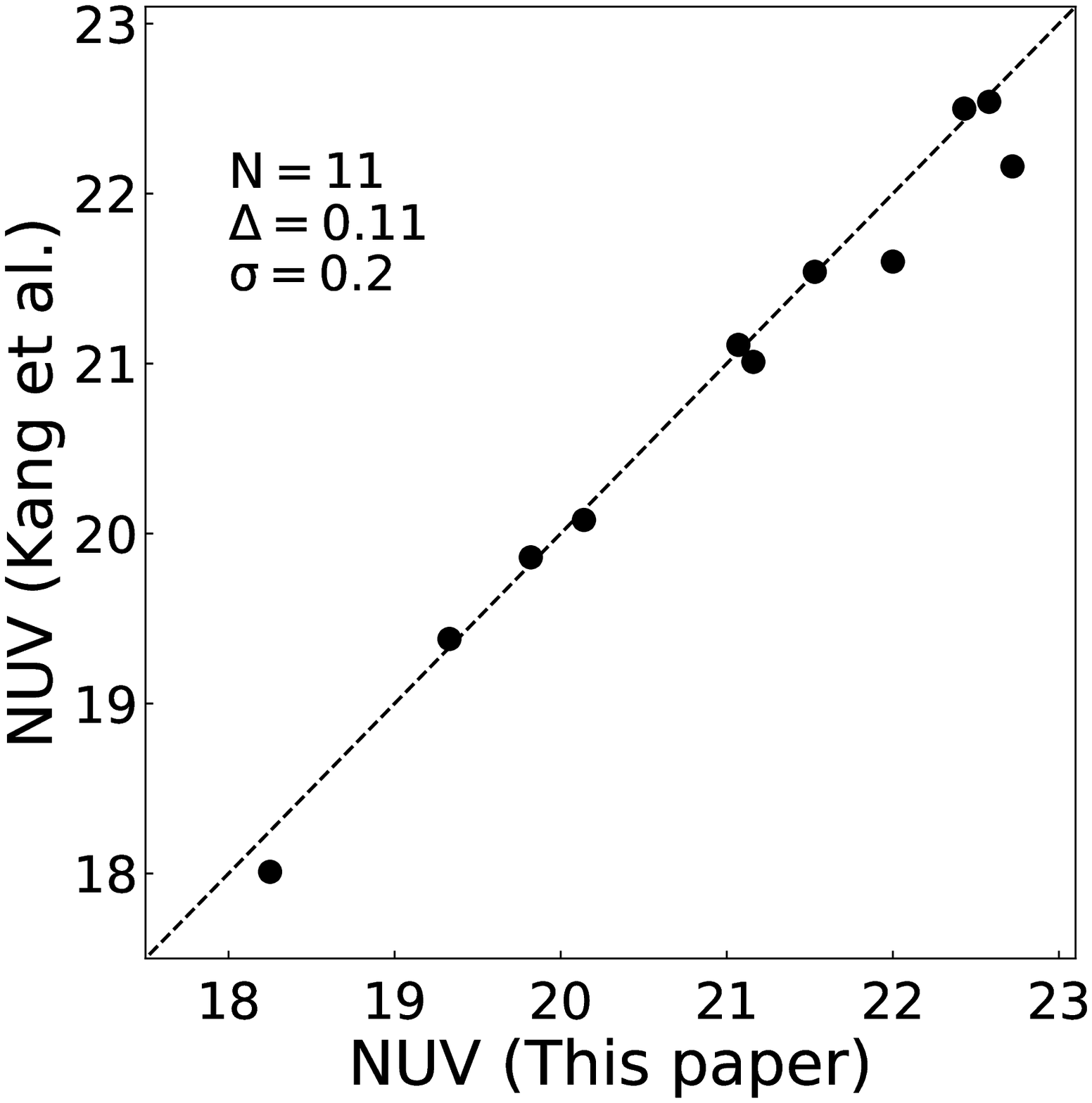}
\caption{Comparison of GALEX FUV (left panel) and NUV (right panel)photometry between \citet{Kang2012} and this paper.}
\label{comkk.fig}
\end{figure}

\begin{figure}
\center
\includegraphics[width=0.24\textwidth]{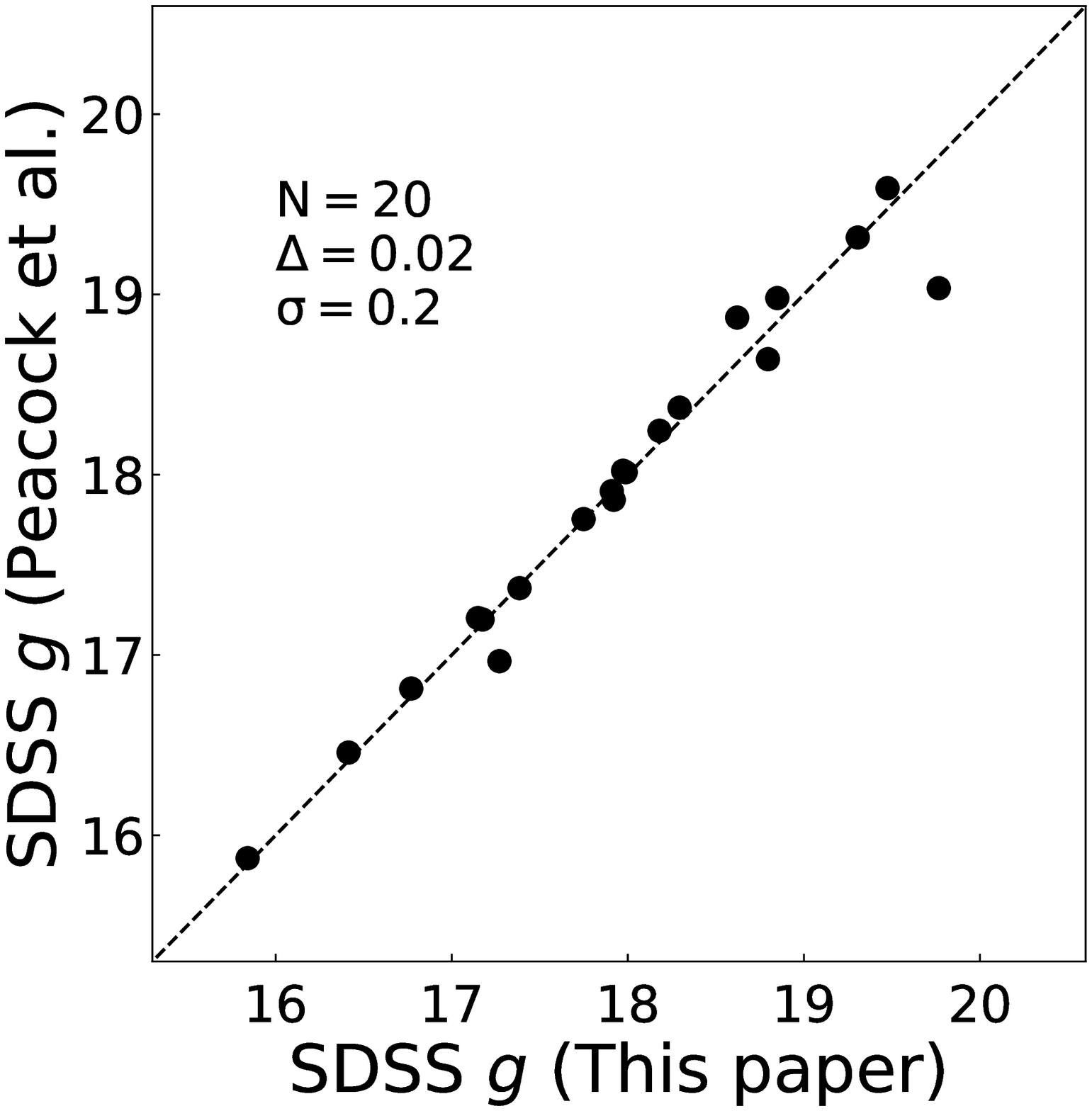}
\includegraphics[width=0.24\textwidth]{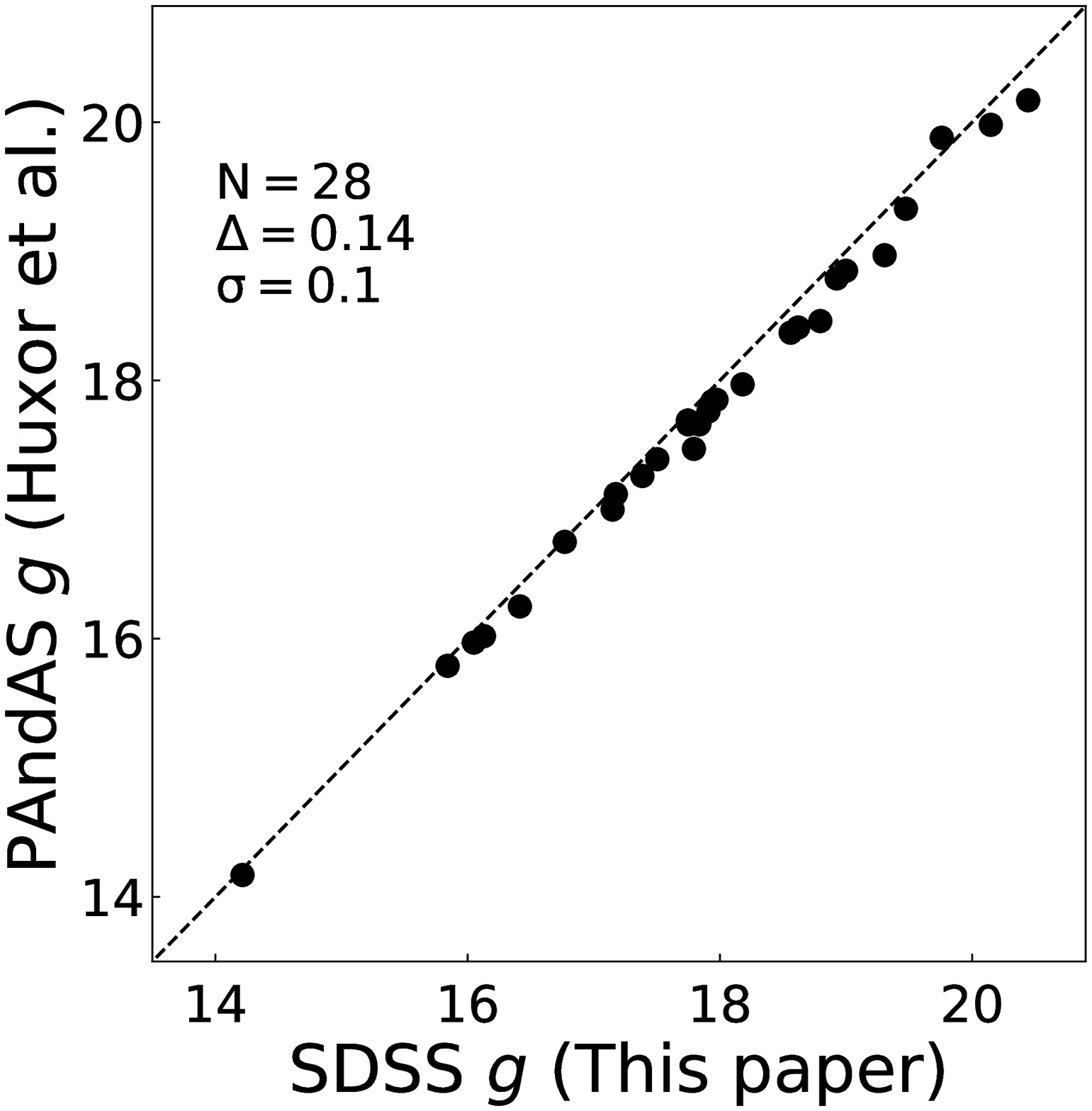}
\caption{Comparison of $g$-band photometry between \citet{Peacock2010} (left panel), \citet{Huxor2014} (right panel) and this paper.}
\label{comph.fig}
\end{figure}

\citet{Kang2012} presented a catalogue of 700 confirmed star clusters in M31, most of which are young disk clusters, and provided the most extensive and updated UV integrated photometry on the AB photometric system based on {\sl GALEX} observations.
Only a few clusters in their catalogue are found in our sample;
the agreement between our NUV photometry and theirs is good (Figure \ref{comkk.fig}).
\citet{Peacock2010} performed $ugriz$ photometry for 1595 M31 clusters and cluster candidates, using the drift scan images of M31 obtained by the SDSS 2.5 m telescope.
\citet{Huxor2014} reported the discovery of 59 halo GCs in M31 from the PAndAS and determined the $g$ and $i$ magnitudes for them.
The photometry from \citet{Peacock2010} are in consistent with our results, but the magnitudes from \citet{Huxor2014} are systematically smaller than ours (Figure \ref{comph.fig}).
This is not surprising since \citet{Huxor2014} used larger radii (compared to ours) for aperture photometry.

We listed the {\sl GALEX} UV, SDSS $ugriz$, and 2MASS NIR photometry of the sample GCs in Table \ref{other.tab}.
The SDSS and 2MASS magnitudes have been transferred to AB magnitudes.
In addition, the magnitudes with an uncertainty larger than 0.3 would not be used in following analysis.
%

\subsection{Reddening and Initial Metallicities}
\label{red.sec}

To robustly and accurately estimate the ages and masses of our sample GCs,
we first determined reddening and metallicities.
For GCs in the outer halo, additional internal M31 reddening towards these objects is likely to be negligible \citep{Huxor2014}.
Therefore, we used the foreground colour excess $E(B-V)$, derived from \citet{Schlafly2011}, as the reddening value of the sample GCs.

Empirically nearly linear relations between different colours and metallicity have been well studied
\citep[e.g.,][]{Barmby2000,Peng2006,Fan2010b,Peacock2011,Harris2017}.
Using the photometry of M31 GCs, \citet{Fan2010b} presented the correlations between the spectroscopic metallicities and 78 BATC colours.
These colours can be used to estimate metallicities of our sample clusters.
For most GCs, we used the colour $(f-o)_{\rm 0}$ --- one of the most metal-sensitive colours --- to estimate the metallicities.
For H8 and HEC9, we used the colour $(f-k)_{\rm 0}$
because we cannot derive accurate $o$-band photometry for these two clusters.
The metallicities will be input as an initial value in the following SED fitting.

\section{METALLICITY, AGE, AND MASS DETERMINATION}

\begin{figure*}[htbp!]
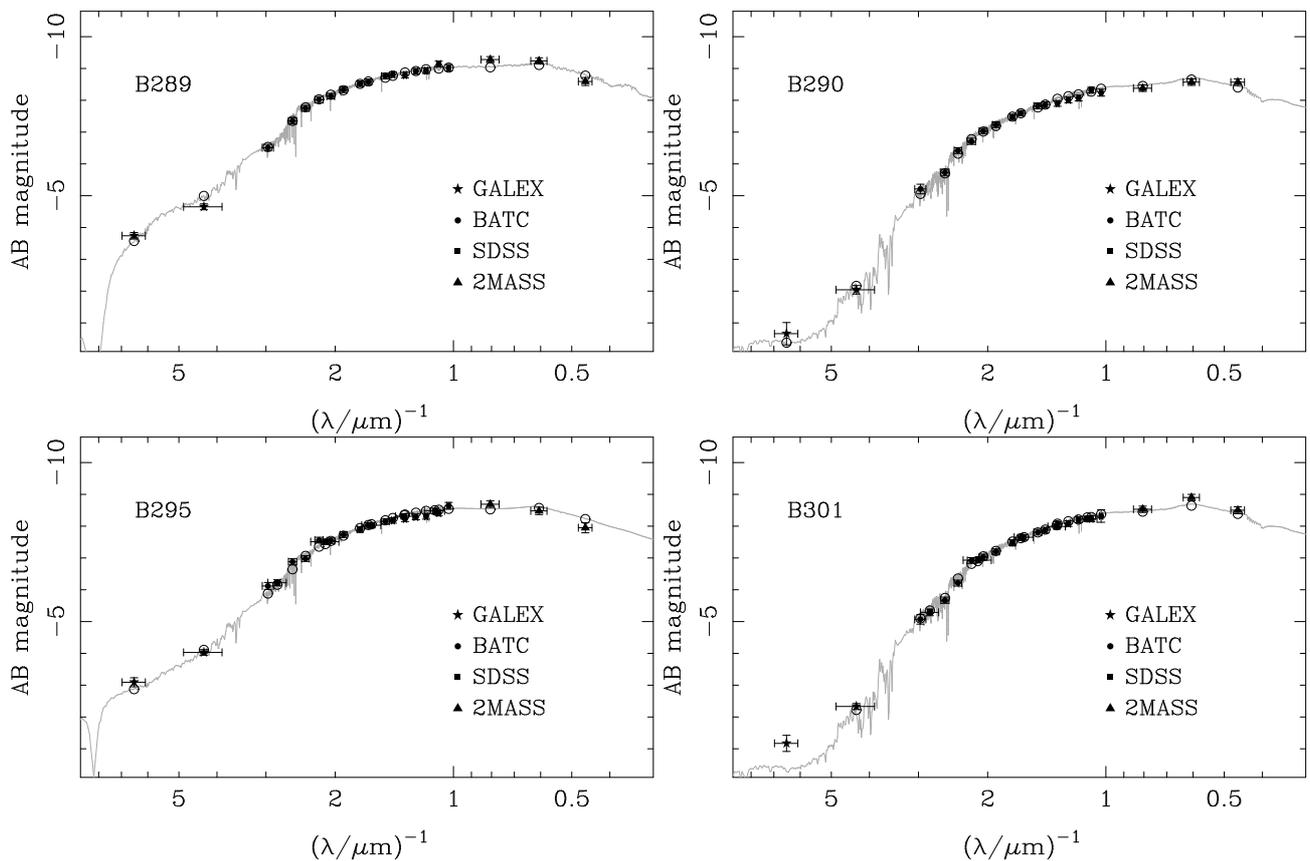

\center
\includegraphics[width=0.46\textwidth]{B289.sed.ps}
\includegraphics[width=0.46\textwidth]{B290.sed.ps}\\
\includegraphics[width=0.46\textwidth]{B295.sed.ps}
\includegraphics[width=0.46\textwidth]{B301.sed.ps}
\caption{Best-fitting integrated SEDs of the BC03 SSP models and the intrinsic SEDs for several example GCs. The filled symbols represent photometric data points, while the open circles represent the calculated magnitudes of the model SEDs for each filter.}
\label{fit.fig}
\end{figure*}

\subsection{Stellar Populations and Synthetic Photometry}

In order to determine the metallicities, ages, and masses of the sample GCs, we compared their SEDs with theoretical stellar population synthesis models.
We adopted the \citet[][hereafter BC03]{Bruzual2003} SSP models,
which are based on the Padova isochrones \citep{Bertelli1994}
and the \citet{Salpeter1955} stellar initial mass function.
The full set of models span the wavelength range from 91 {\AA} to 160 $\mu$m. These models cover ages from 0.1 Myr to 20 Gyr,
and include six initial metallicities, $Z=0.00001, 0.0004, 0.004, 0.008, 0.02$ (solar metallicity), and 0.05,
corresponding to [Fe/H] $= -2.25, -1.65, -0.64, -0.33, 0.09$, and $0.56$.

We convolved the BC03 SSP models with the response curves of the {\sl GALEX} FUV and NUV, SDSS $ugriz$, BATC, and 2MASS $JHK_{\rm s}$ filters to obtain synthetic ultraviolet, optical, and NIR photometry, which would be used to compare with the observational integrated magnitudes.
The synthetic $i{\rm th}$ filter magnitude in the AB magnitude system can be computed as
\begin{equation}
m_i=-2.5\log\frac{\int_{\nu}F_{\nu}\varphi_{i} (\nu)d\nu}{\int_{\nu}\varphi_{i}(\nu)d\nu}-48.60,
\end{equation}
where $F_{\nu}$ is the theoretical SED and $\varphi_{i}$ is the response curve of the $i{\rm th}$ filter of the corresponding photometric systems. Here, $F_{\nu}$ varies with age and metallicity.

\subsection{Fittings}

We used a $\chi^2$ minimization test to determine the SSP model that is most compatible with the observed SED, following
\begin{equation}
\chi^2=\sum_{i=1}^{n}{\frac{[m_{\nu_i}^{\rm intr}-m_{\nu_i}^{\rm mod}(t, Z)]^2}{\sigma_{i}^{2}}},
\end{equation}
where
$m_{\nu_i}^{\rm intr}$ represents the intrinsic integrated magnitude in the $i{\rm th}$ filter,
$m_{\nu_i}^{\rm mod}(t, Z)$ is the integrated magnitude in the same filter of a theoretical SSP at age $t$ and metallicity $Z$,
and $n$ is the number of the filters used for fitting.
The uncertainty $\sigma_i$ is calculated as
\begin{equation}
\sigma_i^{2}=\sigma_{{\rm obs},i}^{2}+\sigma_{{\rm mod},i}^{2}+(R_{\lambda_i}*\sigma_{\rm red})^2
+\sigma_{{\rm md},i}^{2} ,
\end{equation}
which is a combination of the observational uncertainty  $\sigma_{{\rm obs},i}$, the uncertainty associated with the model itself $\sigma_{{\rm mod},i}$, the uncertainty of the reddening value $\sigma_{\rm red}$, and the uncertainty of the distance modulus $\sigma_{{\rm md},i}$.
We adopted $\sigma_{{\rm mod},i}=0.05$ mag for all the filters \citep{Ma2007b,Ma2012,Wang2010}.
The $R_{\lambda_i}$, defined as $A_{\lambda_i}/E(B-V)$, was calculated following \citet{Cardelli1989}.
The $\sigma_{{\rm md},i}$ is always 0.07 mag from
$(m-M)_0 = 24.47\pm0.07$ mag \citep{McConnachie2005}.
During the fitting, the age and metallicity were free parameters,
but the metallicity was constrained in the range (i.e., initial metallicity $\pm$ metallicity uncertainty) derived from the BATC colours (Section \ref{red.sec}).
The upper limit of age in the BC03 models is 20 Gyr, much older than that of the universe. Thus, we adopted a truncated age of 14 Gyr for clusters if their fitted ages are older than that  \citep{Fan2010a,Caldwell2011, Caldwell2016}.
The fitting results of the 53 GCs are listed in Table \ref{agemass.tab}.
As an example, we presented the results of fittings for some sample clusters in Figure \ref{fit.fig}.

\begin{figure}[!h]
\center
\includegraphics[width=0.48\textwidth]{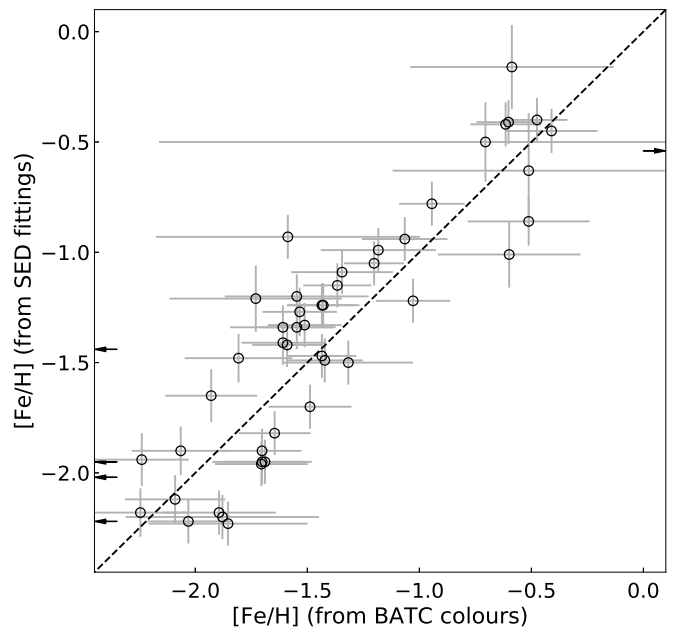}
\caption{Comparison of metallicities deduced from colours and SED fittings.}
\label{fitfeh.fig}
\end{figure}

The comparison of input and fitted metallicities shows good agreement (Figure \ref{fitfeh.fig}).
That means the metallicities deduced from the BATC colours are credible.
One cluster (SK005B) has extremely high metallicity estimation ([Fe/H] $>$ 0.5). The metallicities estimated from other BATC colours (e.g., $(f-k)_0$, $(h-k)_0$) are also higher than 0.5. Then, we estimated the metallicity from other optical and NIR colours (e.g., $(U-R)_0$, $(V-I)_0$, and $(V-K)_0$) using the relations from \citet{Barmby2000}, and all the estimates are higher than 0.5. We suppose that the extinction of SK005B is underestimated. When we used $E(B-V)=0.4$, the values of metallicity derived from the colours are from $-1$ to 0. However, when we fixed the $E(B-V)=0.4$ and re-did the SED fitting, the fitted metallicity is still higher than 0.5. This is unreasonable, and we will remove SK005B in the following analysis.

We determined the masses of the sample GCs sequentially.
The BC03 SSP models are normalized to a total mass of 1 M$_{\odot}$ in stars at age $t=0$; these models provide absolute magnitudes (in the Vega system) in $UBVRI$ and 2MASS $JHK_{\rm s}$ filters. Therefore, the difference between the intrinsic and model absolute magnitudes provides a direct measurement of the cluster mass.
For each GC, we estimated the masses using magnitudes in
the $JHK_{\rm s}$ bands and averaged them as the final cluster mass.
The masses of 13 clusters were not determined, because their $JHK_{\rm s}$ magnitudes were not available here.

\begin{figure*}
\center
\includegraphics[width=0.32\textwidth]{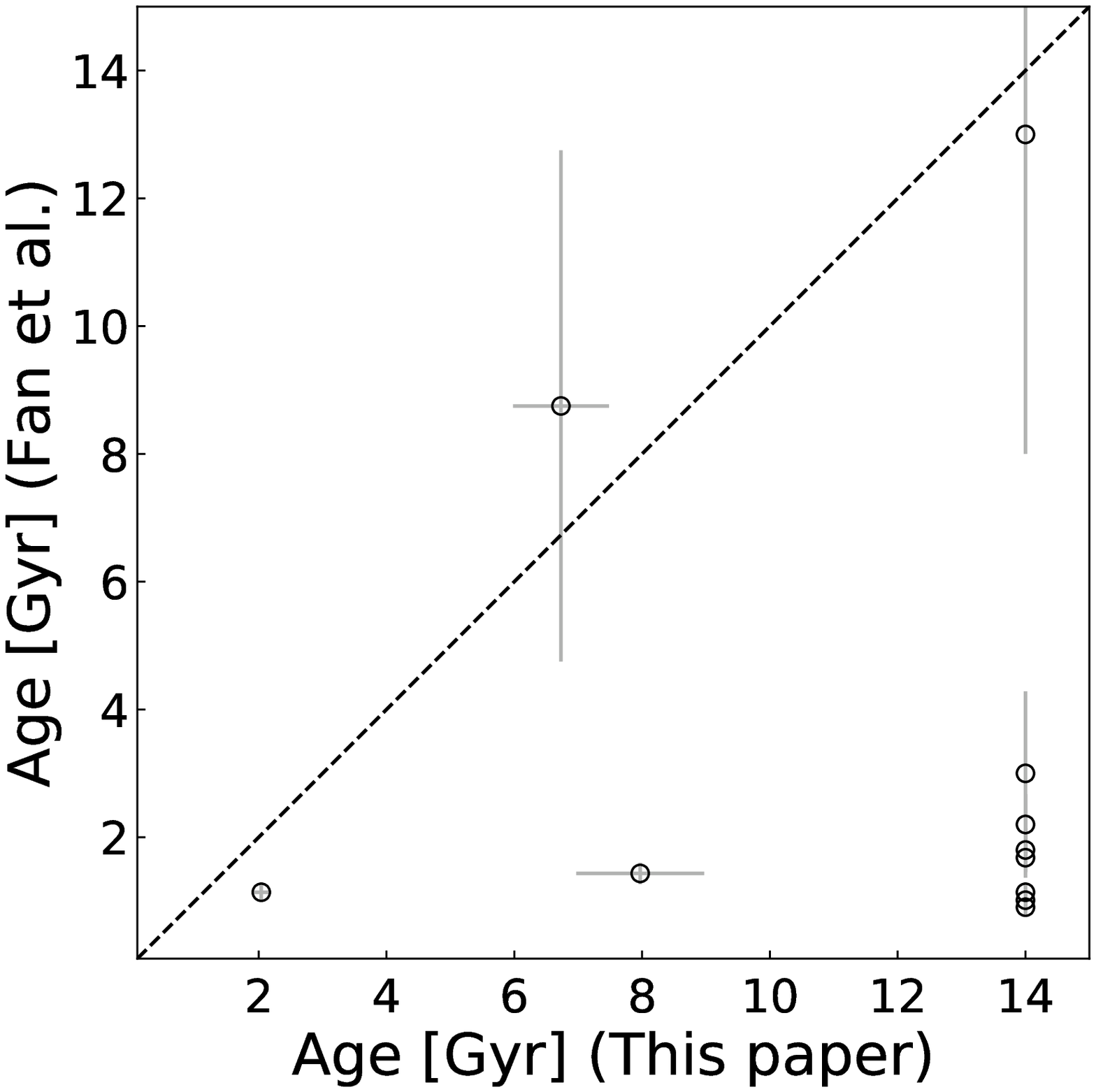}
\includegraphics[width=0.32\textwidth]{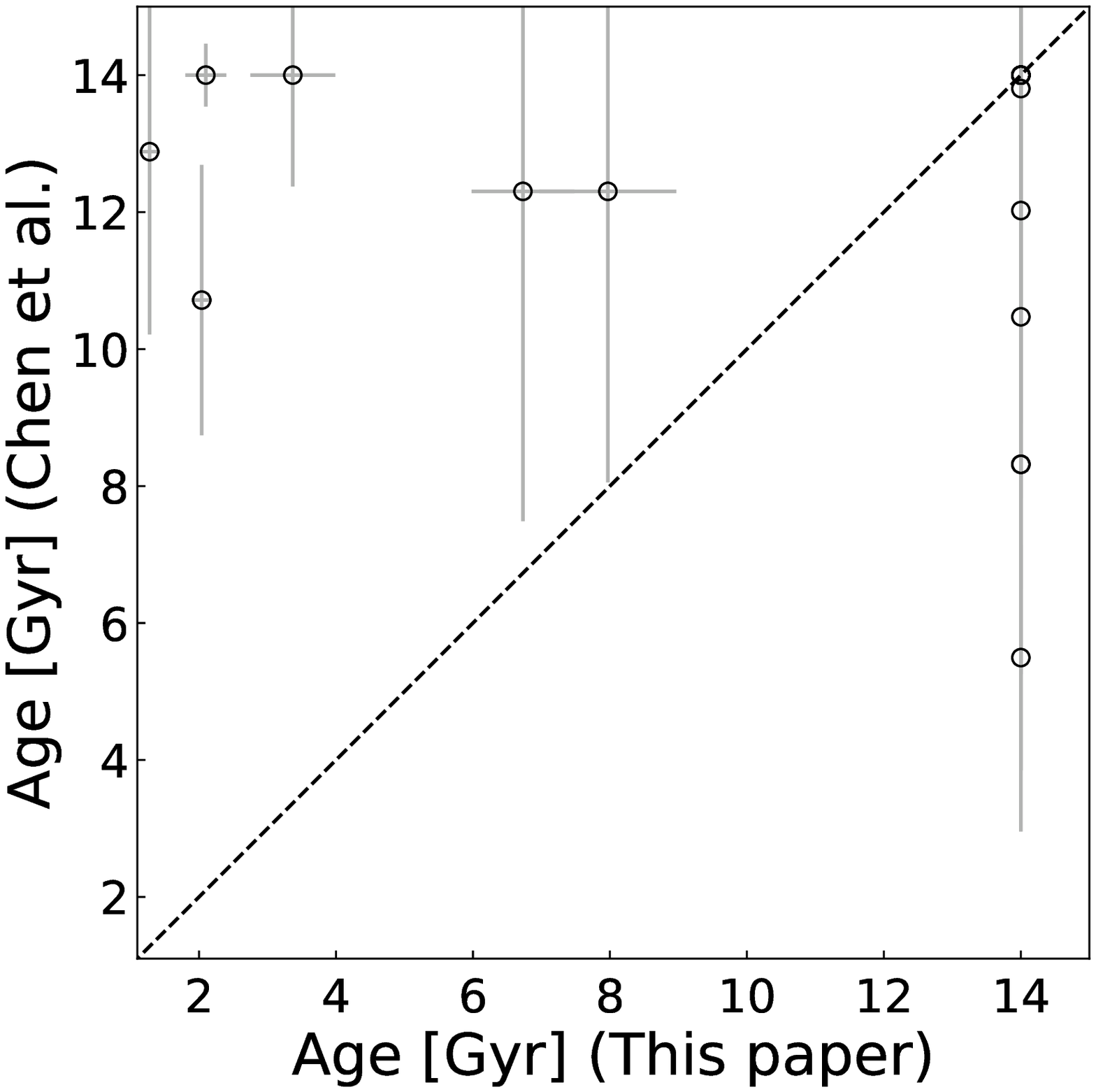}
\caption{Comparison of the ages obtained here with those obtained by previous works (from left panel to right panel): \citet{Fan2010a} and \cite{Chen2016}. For clusters from \cite{Chen2016}, we set their age estimates as 14 Gyr if they are older than that.}
\label{comage.fig}
\end{figure*}

\begin{figure*}
\center
\includegraphics[width=0.32\textwidth]{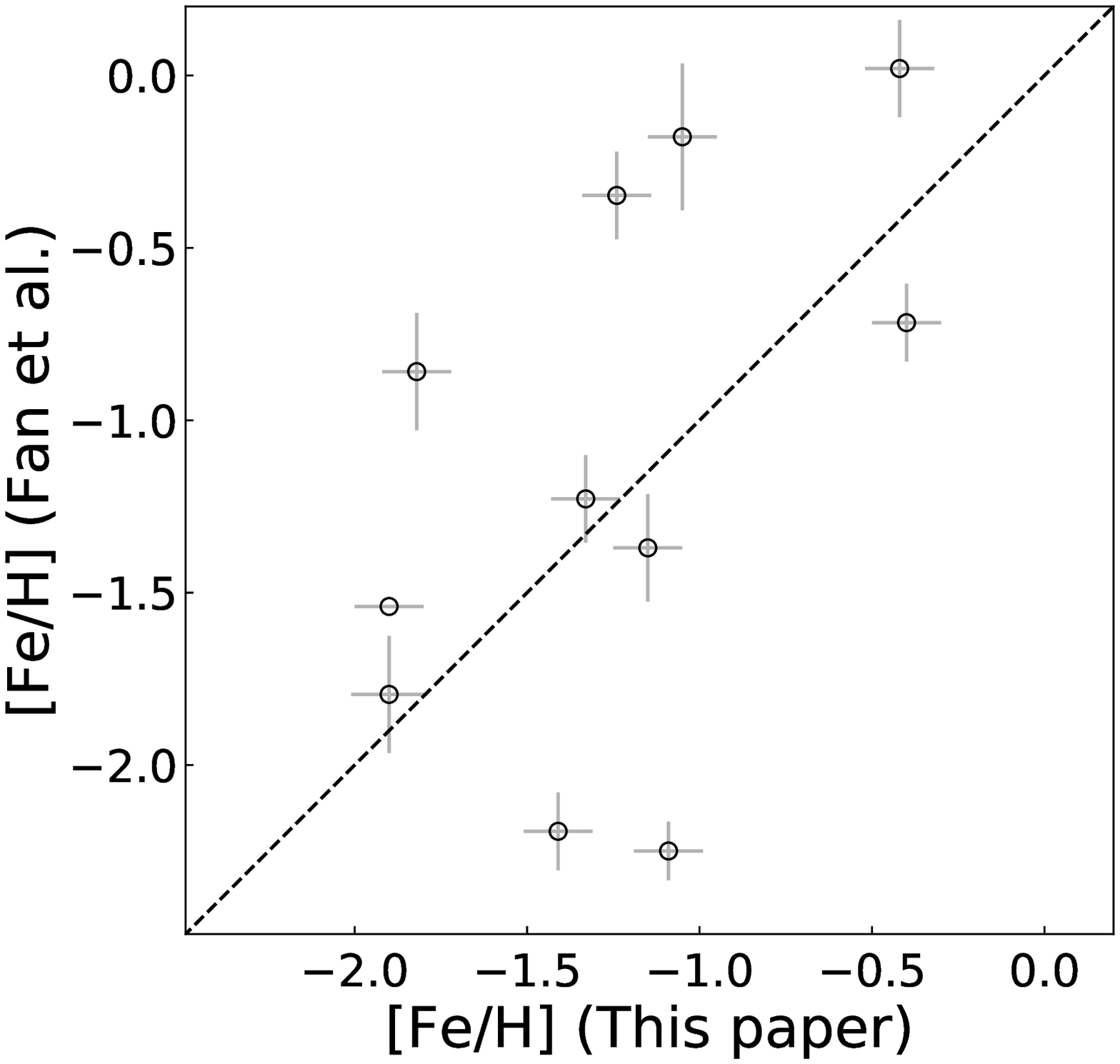}
\includegraphics[width=0.32\textwidth]{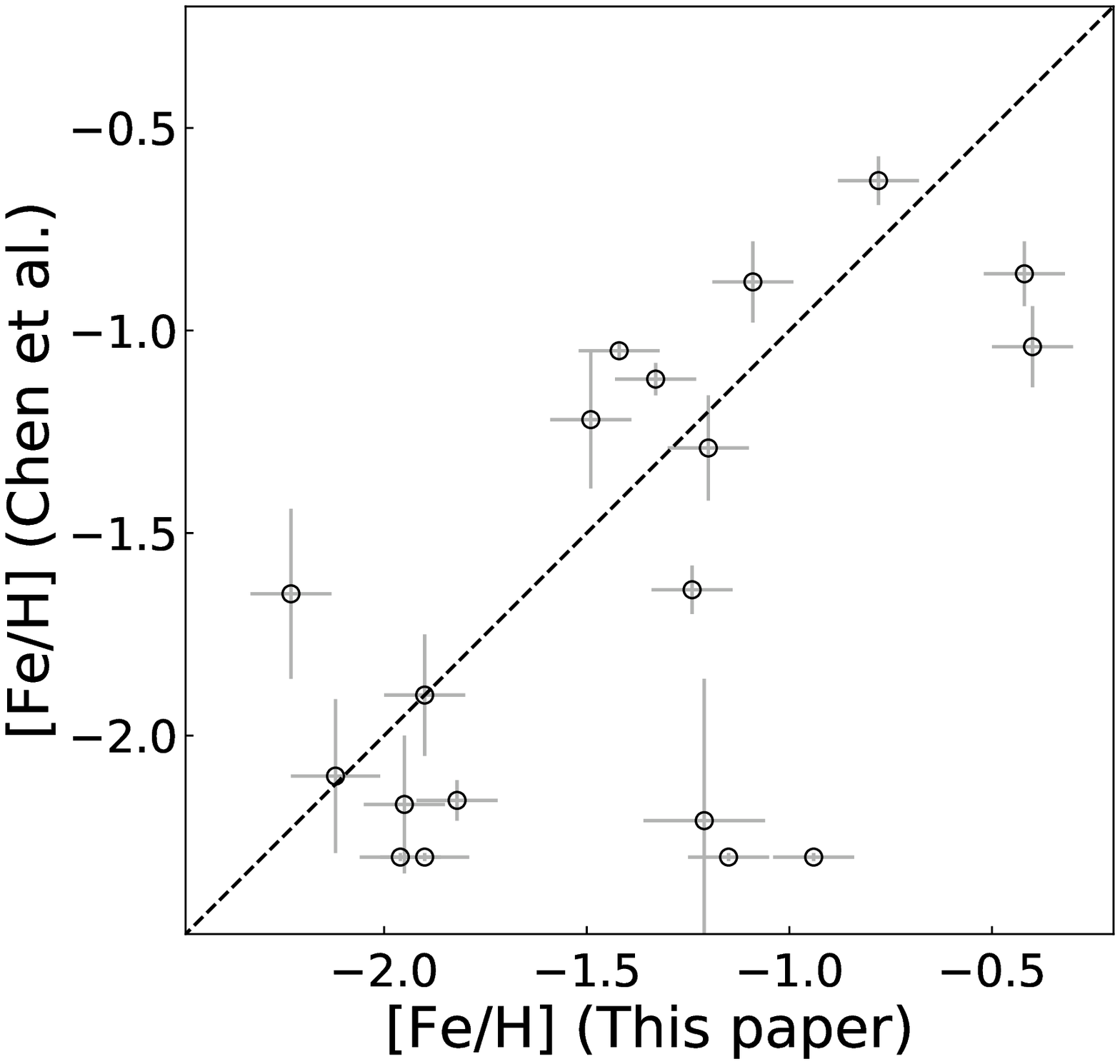}
\includegraphics[width=0.32\textwidth]{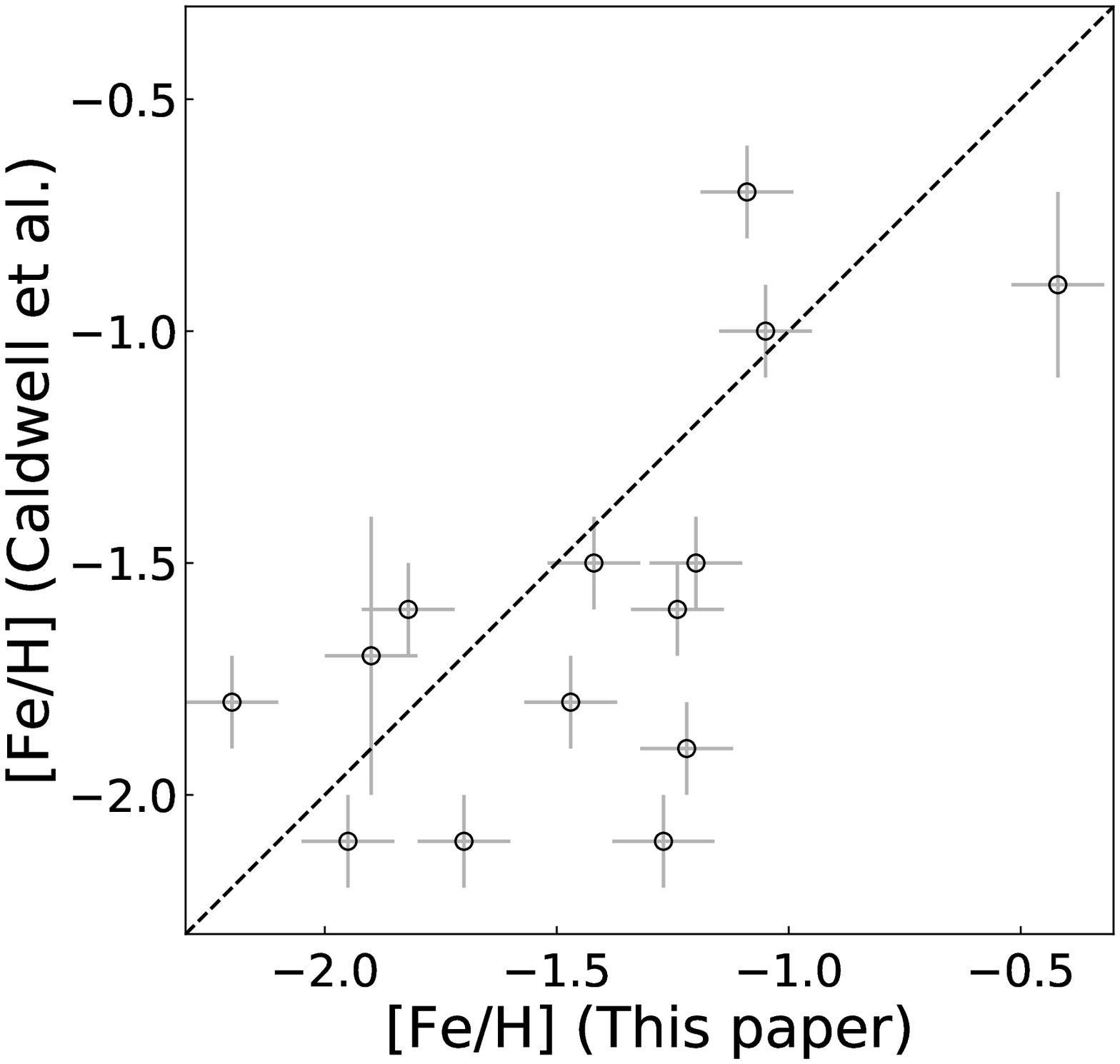}
\caption{Comparison of the metallicities obtained here with those obtained by previous works (from left panel to right panel): \citet{Fan2010a}, \cite{Chen2016}, and \cite{Caldwell2011, Caldwell2016}.}
\label{comfeh.fig}
\end{figure*}

\begin{figure*}
\center
\includegraphics[width=0.32\textwidth]{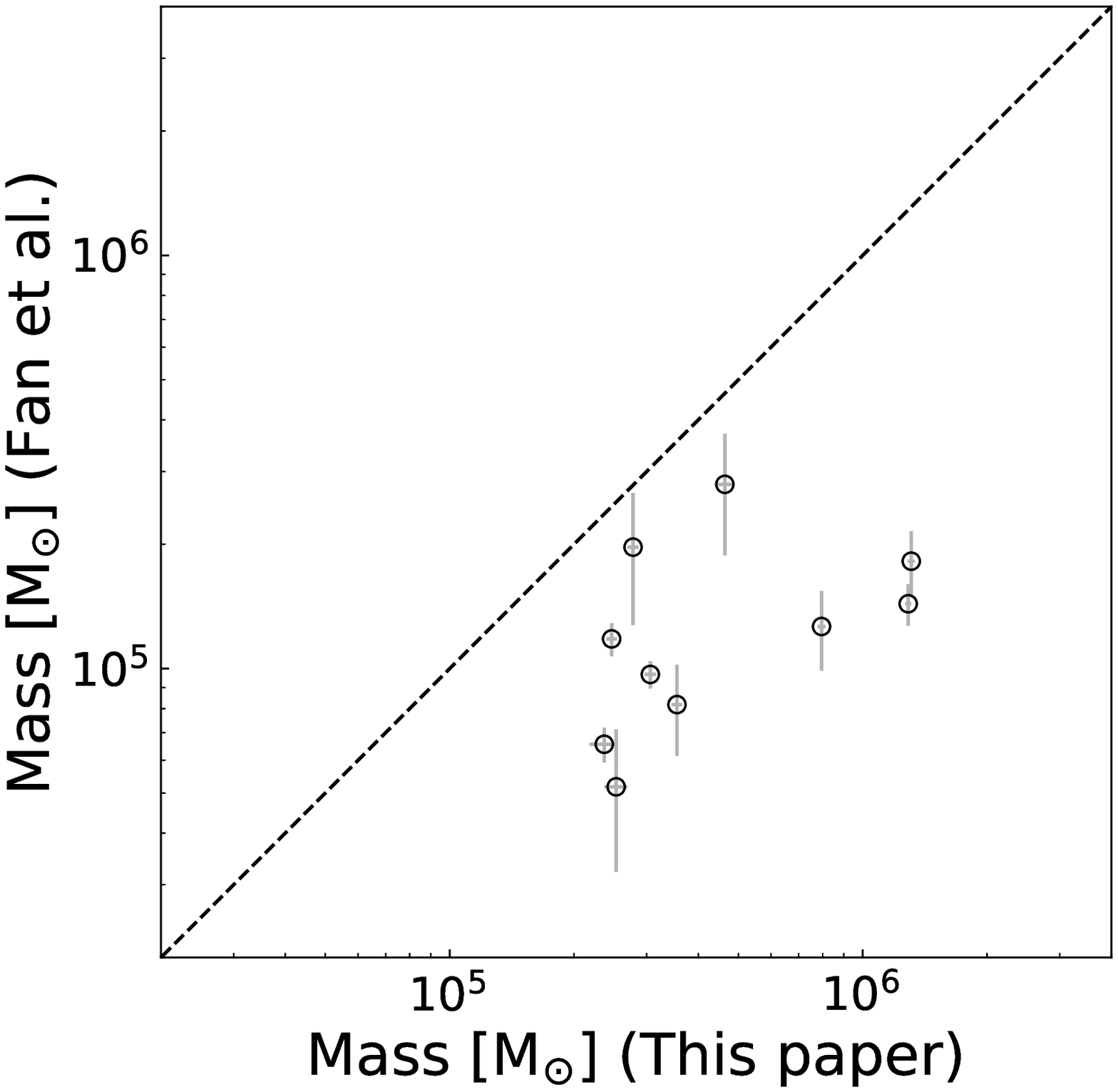}
\includegraphics[width=0.32\textwidth]{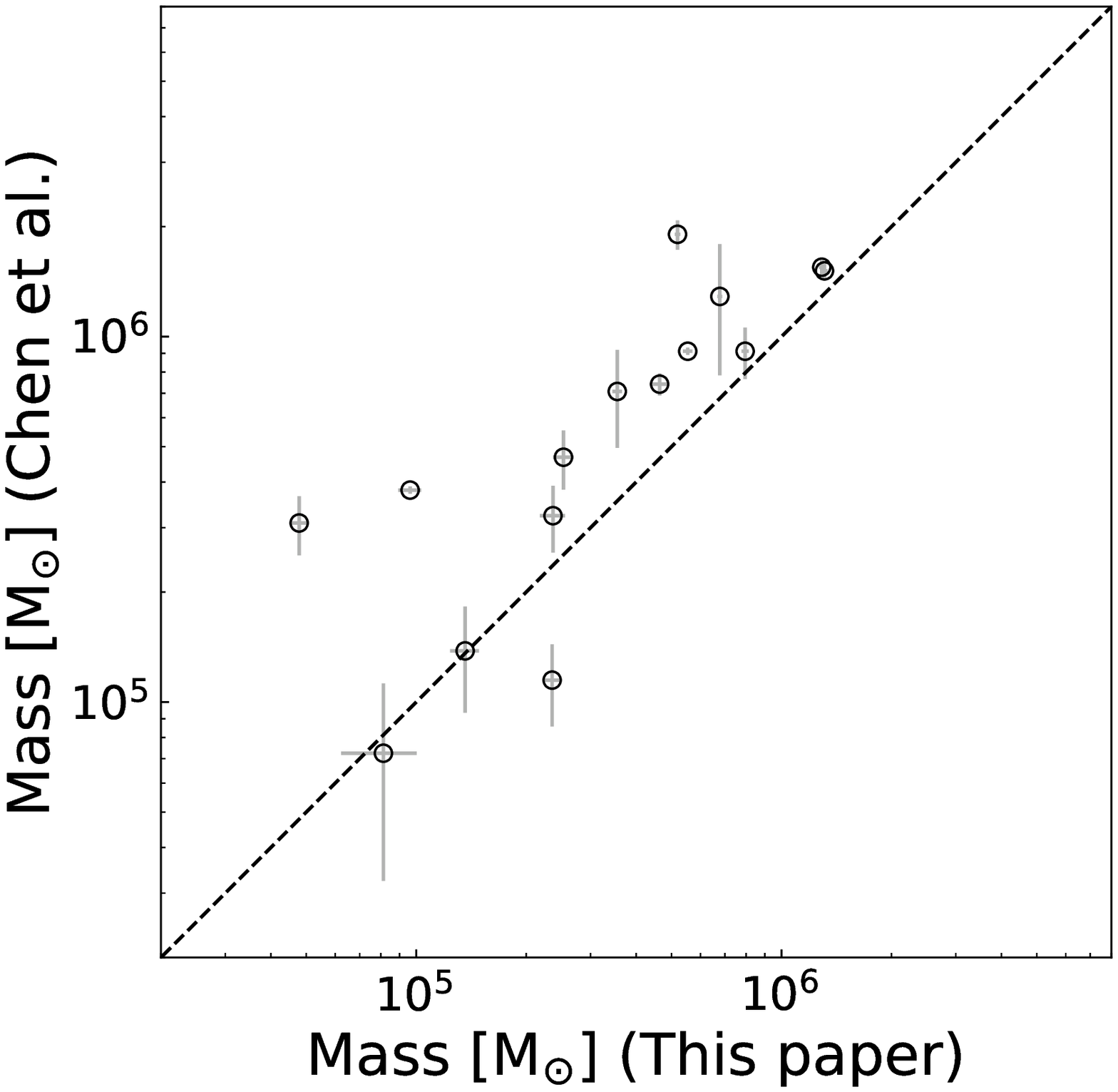}
\includegraphics[width=0.32\textwidth]{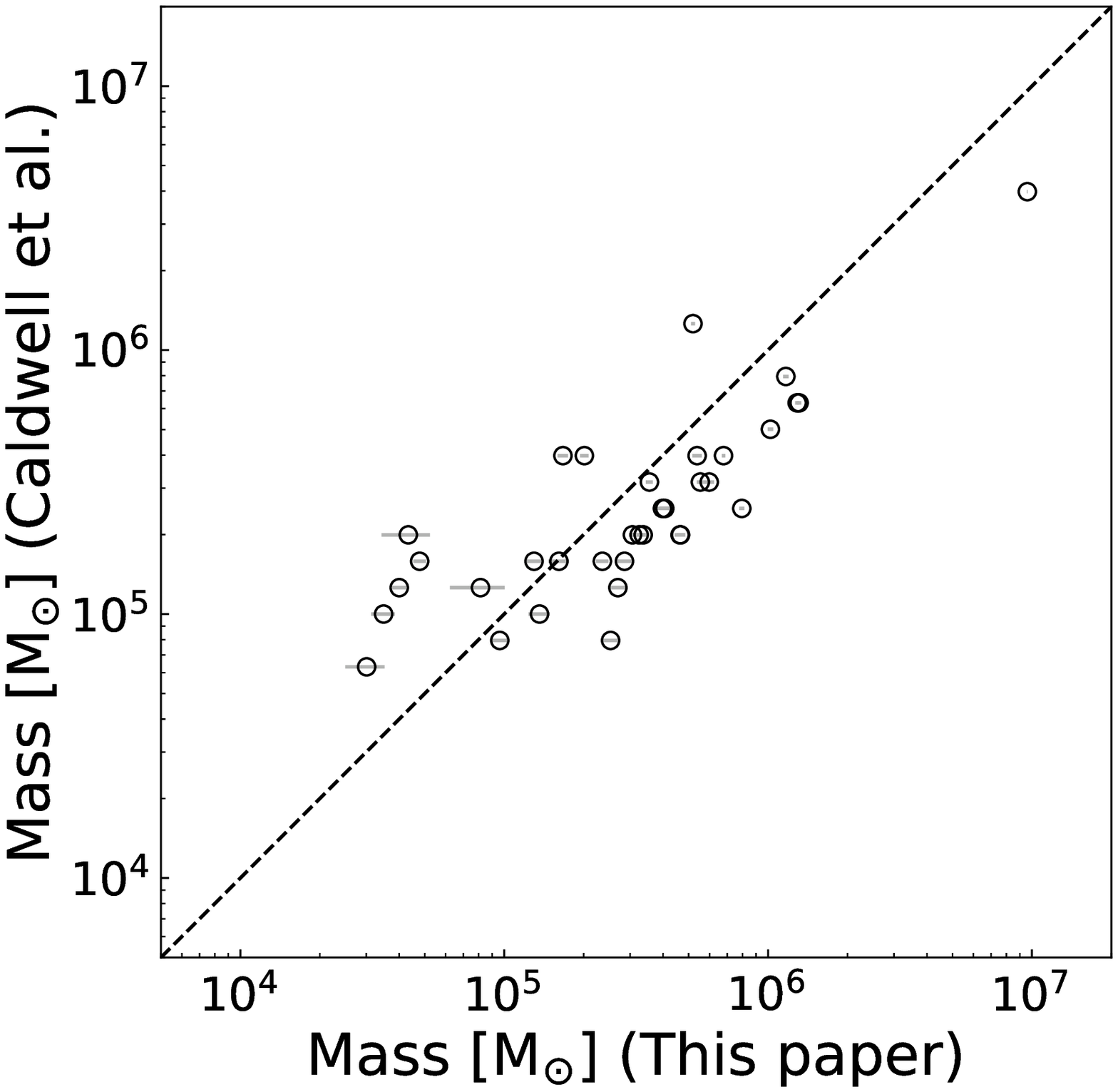}
\caption{Comparison of the masses obtained here with those obtained by previous works (from left panel to right panel): \citet{Fan2010a}, \cite{Chen2016}, and \cite{Caldwell2011, Caldwell2016}.}
\label{commass.fig}
\end{figure*}

\begin{table}[htp!]
\begin{center}
\small
\setlength{\tabcolsep}{0.3em}
\renewcommand\arraystretch{1.1}
\caption{Reddening values, metallicities, ages, and masses of 53 sample GCs in M31.} \vspace{3mm} \label{agemass.tab}
\begin{tabular}{lcccc}
\hline\noalign{\smallskip}
Object &  $E(B-V)$ & [Fe/H] &  Age    &    Mass  \\
       &           &        &  (Gyr)  & ($10^4~{\rm M_\odot}$) \\
    (1)& (2)       &  (3)   &   (4)   &   (5)\\
\hline\noalign{\smallskip}
B289      & 0.07      & $-1.1\pm0.1$         & $2.0\pm0.2$          & $24.7\pm0.7$                   \\
B290      & 0.06      & $-0.4\pm0.1$         & $6.7\pm0.8$          & $27.8\pm0.9$                   \\
B295      & 0.05      & $-1.9\pm0.1$         & 14                  & $46\pm2$                   \\
B301      & 0.05      & $-0.4\pm0.1$         & $8\pm1$          & $36\pm1$                   \\
B400      & 0.08      & $-1.3\pm0.1$         & 14                  & $80\pm2$                   \\
B468      & 0.06      & $-1.1\pm0.1$         & 14                  & $25\pm2$                   \\
B514      & 0.05      & $-1.7\pm0.1$         & 14                  & $117\pm3$                  \\
SDSS-D         & 0.05      & $-2.2\pm0.1$         & 14                  & $21\pm1$                   \\
EXT8      & 0.06      & $-2.0\pm0.1$         & $3.4\pm0.6$          & $51.9\pm0.9$                   \\
G001      & 0.05      & $-1.1\pm0.1$         & 14                  & $961\pm3$                  \\
G002      & 0.05      & $-1.8\pm0.1$         & 14                  & $131\pm3$                  \\
G268      & 0.07      & $-0.8\pm0.1$         & 14                  & $68\pm1$                   \\
G327      & 0.07      & $-1.2\pm0.1$         & 14                  & $129\pm2$                  \\
G339      & 0.08      & $-1.4\pm0.1$         & 14                  & $31\pm1$                   \\
G353      & 0.07      & $-1.9\pm0.1$         & 14                  & $24\pm2$                   \\
H2        & 0.05      & $-2.2\pm0.1$         & $13\pm3$         & $27\pm2$                   \\
H4        & 0.06      & $-1.2\pm0.1$         & 14                  & $40\pm2$                   \\
H5        & 0.07      & $-1.3\pm0.1$         & $2.0\pm0.2$          & $20.1\pm0.8$                   \\
H7        & 0.05      & $-2.1\pm0.1$         & 14                  & $9.6\pm0.7$                    \\
H8        & 0.05      & $-0.5\pm0.2$         & $9\pm4$          & ...                           \\
H9        & 0.05      & $-1.0\pm0.1$         & $1.3\pm0.2$          & $4.0\pm0.3$                    \\
H10       & 0.06      & $-1.5\pm0.1$         & 14                  & $102\pm2$                  \\
H11       & 0.06      & $-2.2\pm0.1$         & 14                  & $41\pm2$                   \\
H12       & 0.06      & $-1.2\pm0.1$         & $2.2\pm0.2$          & $16.7\pm0.8$                   \\
H15       & 0.05      & $-1.3\pm0.1$         & $4\pm1$          & $3.5\pm0.4$                    \\
H16       & 0.07      & $-1.5\pm0.1$         & 14                  & $29\pm2$                   \\
H17       & 0.04      & $-1.5\pm0.1$         & 14                  & $24\pm1$                   \\
H18       & 0.08      & $-1.9\pm0.1$         & 14                  & $60\pm3$                   \\
H19       & 0.05      & $-0.9\pm0.1$         & $1.3\pm0.2$          & $4.8\pm0.3$                    \\
H20       & 0.05      & $-1.2\pm0.2$         & $2.1\pm0.3$          & ...                           \\
H22       & 0.04      & $-1.9\pm0.1$         & 14                  & $34\pm2$                   \\
H23       & 0.04      & $-1.4\pm0.1$         & 14                  & $55\pm2$                   \\
H24       & 0.09      & $-1.2\pm0.1$         & 14                  & $14\pm1$                   \\
H25       & 0.08      & $-1.3\pm0.1$         & $2.6\pm0.3$          & $13\pm1$                   \\
H27       & 0.05      & $-1.9\pm0.1$         & 14                  & $54\pm2$                   \\
HEC7      & 0.08      & $-2.2\pm0.1$         & 14                  & $4.3\pm0.9$                    \\
HEC9      & 0.05      & $-2.0\pm0.5$         & 14                  & ...                           \\
HEC10     & 0.09      & $-1.4\pm0.6$         & 14                  & ...                           \\
HEC11     & 0.04      & $-2.2\pm0.1$         & 14                  & $8\pm2$                    \\
HEC13     & 0.04      & $-2.2\pm0.6$         & 14                  & ...                           \\
PAndAS-19 & 0.05      & $-0.6\pm0.3$         & $3\pm1$          & ...                           \\
PAndAS-20 & 0.06      & $-0.2\pm0.2$         & $2.8\pm0.6$          & ...                           \\
PAndAS-21 & 0.05      & $-1.6\pm0.1$         & $9\pm2$          & $16\pm1$                   \\
PAndAS-22 & 0.06      & $-0.9\pm0.1$         & $5\pm1$          & $3.0\pm0.5$                    \\
PAndAS-23 & 0.05      & $-0.5\pm0.2$         & $7\pm2$          & ...                           \\
PAndAS-25 & 0.06      & $-2.0\pm0.4$         & 14                  & ...                           \\
PAndAS-26 & 0.05      & $-0.9\pm0.1$         & 14                  & $2.1\pm0.4$                    \\
PAndAS-34 & 0.06      & $-0.5\pm0.1$         & $6.4\pm0.6$          & $12.5\pm0.9$                   \\
PAndAS-36 & 0.06      & $-1.9\pm0.1$         & 14                  & $33\pm2$                   \\
PAndAS-37 & 0.05      & $-0.4\pm0.1$         & 14                  & $47\pm2$                   \\
PAndAS-39 & 0.08      & $-1.0\pm0.2$         & $6\pm2$          & ...                           \\
PAndAS-41 & 0.08      & $-1.5\pm0.1$         & 14                  & $30\pm2$                   \\
SK005B    & 0.06      & $0.5\pm0.1$          & $13\pm3$         & $20.0\pm0.9$                   \\
\hline
\end{tabular}
\end{center}
\end{table}

\subsection{Comparison with Previous Determinations}

In this paper, we determined the metallicities, ages, and masses for 53 GCs by comparing their multicolour photometries with theoretical SSP models.
In Figure \ref{comage.fig}, \ref{comfeh.fig}, and \ref{commass.fig}, we compared our estimates with those in the literature: \citet{Fan2010a}, \citet{Chen2016}, and \citet{Caldwell2011, Caldwell2016}.

\citet{Fan2010a} presented updated $UBVRI$ photometry for 970 star clusters in M31, and determined the metallicities, ages, and masses by comparing the integrated SEDs ($UBVRI$) with BC03 SSP models.
Eleven of them are found in our sample.
There are some discrepancy between our metallicity estimates and theirs,
and their age and mass estimates are systematically smaller than ours.
This is likely caused by the larger reddening values they used. Of these 11 shared clusters, the three ones (B289, G339, and G353) with the largest $E(B-V)$ values from 0.2 to 0.5, obtained from free fitting \citep{Fan2010a}, are also the youngest. Considering that the reddening, metallicity, and age are degenerate during the SED fitting, their fitting results may be inaccurate.
Furthermore, our SEDs includes the photometric data in many bands, especially in the $UV$ bands,
which can help derive accurate ages of star clusters \citep{deGrijs2003, Anders2004}.

\citet{Caldwell2011} estimated metallicities using the Lick <Fe> indices, and determined ages for clusters with [Fe/H] $>$ $-1$ using the automatic stellar population analysis programme EZ\_Ages.
\citet{Caldwell2016} collected metallicities for a number of M31 GCs from previous studies \citep[e.g.,][]{Mackey2006,Mackey2007,Colucci2014}.
Forty-three objects are in common between our sample and those of \citet{Caldwell2011, Caldwell2016}.
Both their metallicities and mass estimates are in agreement with ours.
\citet{Caldwell2011, Caldwell2016} reported an age of 14 Gyr for all these clusters, since they simply assigned an age of 14 Gyr to all clusters with [Fe/H] $< -0.95$ dex,
meaning that no precise age was determined.
In our study, among these sources,
15 objects have ages younger than 14 Gyr (i.e.,
four objects have ages between 5 and 14 Gyr,
and 11 objects have ages ranging from 1 to 5 Gyr).

Using the LAMOST spectra of M31 star clusters,
\citet{Chen2016} determined metallicities, ages, and masses from full spectral fitting with the PEGASE-HR models.
Nineteen of them are included in our sample.
For metallicity and mass, the agreement between our estimates and their estimates is generally good.
The discrepancy between the age estimates is clear, but the uncertainties of the ages are quite large.
Four clusters (B289, EXT8, H19, and H20) were estimated to be older than 10 Gyr in \citet{Chen2016}, but younger than 4 Gyr (i.e., 2.0, 3.4, 1.3, and 2.1 Gyr) in our study.
For H19, the photometry is not good enough for accurate SED fitting because it is too faint.
We checked the LAMOST observations of the four clusters, and found the S/Ns of all these spectra are quite low:
the S/Ns are ranging from 2 to 10 in the $g$ band and ranging from 3 to 17 in the $r$ band.
Therefore, the age estimates from spectral fitting may be inaccurate for these clusters.

We also collected metallicities from several other studies.
Using the Keck/HIRES spectra,
\citet{Alves-Brito2009} and \citet{Colucci2014} measured the metallicities for a small sample of M31 GCs.
\citet{Sakari2015} presented detailed chemical abundances for seven M31 outer halo GCs, which were derived from the spectra taken with the Hobby-Eberly Telescope.
Based on the {\it HST} observations, the metallicities for a group of M31 GCs were measured with the colour-magnitude diagram (CMD) fitting \citep{Rich2005,Mackey2006,Mackey2007}.
Although the results from different methods may be discrepant,
in general, most of our estimates are consistent with those derived from high-resolution spectra or CMD fitting (Table \ref{fehcom.tab}).

\begin{table*}[htp!]
\begin{center}
\small
\setlength{\tabcolsep}{0.3em}
\renewcommand\arraystretch{1.05}
\caption{Metallicities for GCs measured from this paper and previous studies.} \vspace{3mm} \label{fehcom.tab}
\begin{tabular}{lcccccccccc}
\hline\noalign{\smallskip}
\multirow{2}{*}{Object} &   \multicolumn{10}{c}{[Fe/H]}   \\
\cmidrule(l){2-11}
 & this paper & F10 & Ch17 & A09 & Ca11 & S15 & R05 & M06 & M07 & Co14  \\
    (1)& (2)       &  (3)   &   (4)  & (5)& (6)       &  (7)   &   (8) & (9)& (10)       &    (11) \\
\hline\noalign{\smallskip}
B289  &  $-1.1\pm0.1$  &  $-1.37\pm0.16$ &$-2.30\pm0.01$ & ...& ...& ...& ...& ...& ...& ...\\
B290  &  $-0.4\pm0.1$  &  $-0.72\pm0.11$ &$-1.04\pm0.10$ & ...& ...& ...& ...& ...& ...& ...\\
B295  &  $-1.9\pm0.1$  &  $-1.54\pm0.00$ &$-1.90\pm0.15$ & ...&$-1.7\pm0.3$ & ...& ...& ...& ...& ...\\
B301  &  $-0.4\pm0.1$  &  $0.02\pm0.14$ &$-0.86\pm0.08$ & ...&$-0.9\pm0.2$ & ...& ...& ...& ...& ...\\
B400  &  $-1.3\pm0.1$  &  $-1.23\pm0.13$ &$-1.12\pm0.04$ & ...& ...& ...& ...& ...& ...& ...\\
B468  &  $-1.1\pm0.1$  &  $-2.25\pm0.09$ &$-0.88\pm0.10$ & ...& ...& ...&$-0.7$ & ...& ...& ...\\
B514  &  $-1.7\pm0.1$  &   ...& ...& ...& ...& ...& ...& ...&$-2.14$ &$-1.74$ \\
EXT8  &  $-2.0\pm0.1$  &   ...&$-2.30\pm0.01$ & ...& ...& ...& ...& ...& ...& ...\\
G001  &  $-1.1\pm0.1$  &  $-0.18\pm0.21$ & ...& ...& ...& ...&$-1.0$ & ...& ...& ...\\
G002  &  $-1.8\pm0.1$  &  $-0.86\pm0.17$ &$-2.16\pm0.05$ & ...& ...& ...& ...& ...& ...&$-1.63$ \\
G268  &  $-0.8\pm0.1$  &   ...&$-0.63\pm0.06$ & ...& ...& ...& ...& ...& ...& ...\\
G327  &  $-1.2\pm0.1$  &  $-0.35\pm0.13$ &$-1.64\pm0.06$ & ...& ...& ...& ...& ...& ...&$-1.65$ \\
G339  &  $-1.4\pm0.1$  &  $-2.19\pm0.11$ & ...& ...& ...& ...& ...& ...& ...& ...\\
G353  &  $-1.9\pm0.1$  &  $-1.79\pm0.17$ &$-2.30\pm0.01$ & ...& ...& ...& ...& ...& ...& ...\\
H4  &  $-1.2\pm0.1$  &   ...& ...& ...& ...& ...& ...& ...&$-1.94$ & ...\\
H5  &  $-1.3\pm0.1$  &   ...& ...& ...& ...& ...& ...& ...&$-2.14$ & ...\\
H7  &  $-2.1\pm0.1$  &   ...&$-2.10\pm0.19$ & ...& ...& ...& ...& ...& ...& ...\\
H10  &  $-1.5\pm0.1$  &   ...& ...&$-1.40\pm0.10$ & ...&$-1.36\pm0.02$ & ...& ...&$-1.84$ &$-1.45$ \\
H17  &  $-1.5\pm0.1$  &   ...&$-1.22\pm0.17$ & ...& ...& ...& ...& ...& ...& ...\\
H19  &  $-0.9\pm0.1$  &   ...&$-2.30\pm0.01$ & ...& ...& ...& ...& ...& ...& ...\\
H20  &  $-1.2\pm0.2$  &   ...&$-2.21\pm0.35$ & ...& ...& ...& ...& ...& ...& ...\\
H23  &  $-1.4\pm0.1$  &   ...&$-1.05\pm0.02$ & ...& ...&$-1.12\pm0.02$ & ...& ...&$-1.54$ & ...\\
H24  &  $-1.2\pm0.1$  &   ...&$-1.29\pm0.13$ & ...& ...& ...& ...& ...&$-1.54$ & ...\\
H27  &  $-1.9\pm0.1$  &   ...&$-2.17\pm0.17$ &$-1.73\pm0.10$ & ...& ...& ...& ...&$-2.14$ & ...\\
HEC7  &  $-2.2\pm0.1$  &   ...& ...& ...& ...& ...& ...&$-1.84$ & ...& ...\\
HEC11  &  $-2.2\pm0.1$  &   ...&$-1.65\pm0.21$ & ...& ...& ...& ...& ...& ...& ...\\
\hline
\end{tabular}
\end{center}
{Note. References for metallicities in previous studies:
R05 --- \citet{Rich2005};
M06 --- \citet{Mackey2006};
M07 --- \citet{Mackey2007};
A09 --- \citet{Alves-Brito2009};
F10 --- \citet{Fan2010a};
Ca11 --- \citet{Caldwell2011};
Co14 --- \citet{Colucci2014};
S15 --- \citet{Sakari2015};
Ch16 --- \citet{Chen2016}.}
\end{table*}

\section{DISCUSSION}

\subsection{Spatial Distribution}

Using these halo GCs with homogeneously determined metallicities, ages, and masses, we can now investigate their spatial distribution.
To derive the de-projected galactocentric distance,
firstly,  we used an $X,Y$ projection to refer to the relative positions of the objects \citep{Huchra1991, Perrett2002}.
The $X$ coordinate projects along M31's major axis, increasing towards the northeast;
the $Y$ coordinate extends along the minor axis of the M31 disk,
where positive $Y$ increases towards the northwest.
We adopted $\alpha_0=00^{\rm h}42^{\rm m}44^{\rm s}.30$ and $\delta_0=+41^\circ16'09''.0$ (J2000.0) as the certre of M31,
and used a position angle of PA $=38^\circ$ for the major axis \citep{ken89}.
Formally,
 \begin{equation}
\begin{split}
&X = A\sin{\rm (PA)} + B\cos{\rm (PA)} \ \ \ {\rm and}  \\
&Y = -A\cos{\rm (PA)} + B\sin{\rm (PA)} ,
\end{split}
\end{equation}
where $A=\sin(\alpha-\alpha_0)\cos\delta$ and $B=\sin\delta \cos\delta_0 - \cos(\alpha-\alpha_0)
\cos\delta \sin\delta_0$.
Then, using an inclination angle of IA $=77.5^\circ$ for the M31's disk, we calculated the de-projected radius as
$\sqrt{X^2+(Y/\cos{\rm (IA)})^2}$.

\begin{figure}
\center
\includegraphics[width=0.49\textwidth]{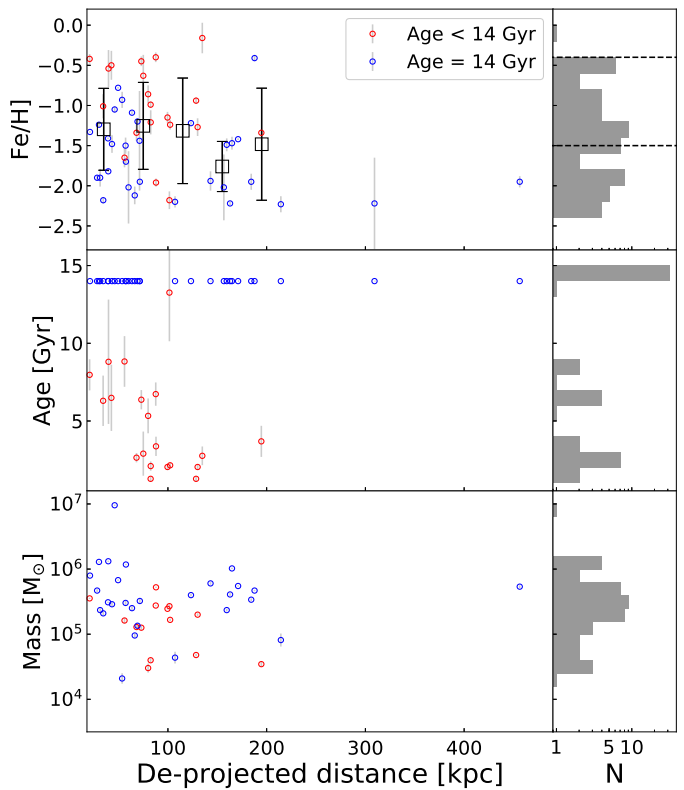}
\caption{Top panel: distribution of [Fe/H] of GCs against de-projected radius. The red and blue circles represent the GCs younger and older than 14 Gyr, respectively.
The squares represent the mean metallicity values in different de-projected galactocentric distance for the GCs.
The two dashed lines represent metallicities [Fe/H] $=$ $-$0.4 and [Fe/H] $=$ $-$1.5. Middle panel: age vs. de-projected radius. Bottom panel: mass vs. de-projected radius.}
\label{project.fig}
\end{figure}

Figure \ref{project.fig} shows the metallicity, ages, and masses of the sample clusters as a function of de-projected radius.
%
The metallicity shows no clear relation with the de-projected galactocentric distance within 200 kpc. Most of the GCs (51 out of 53) have metallicities lower than $-$0.4, and can be divided into one intermediate-metallicity group and one metal-poor group \citep{Caldwell2016}.
The clusters with age younger than $\approx$ 8 Gyr are mostly located at de-projected distances around 100 kpc, but this may be caused by a selection effect.
Most GCs have masses between 10$^4$ -- 10$^6$ M$_{\odot}$,
except that the most massive cluster G001 has a mass around 10$^7$ M$_{\odot}$.
No clear relation is seen for the mass and de-projected distance.

\begin{figure}[!ht]
\center
\includegraphics[width=0.47\textwidth]{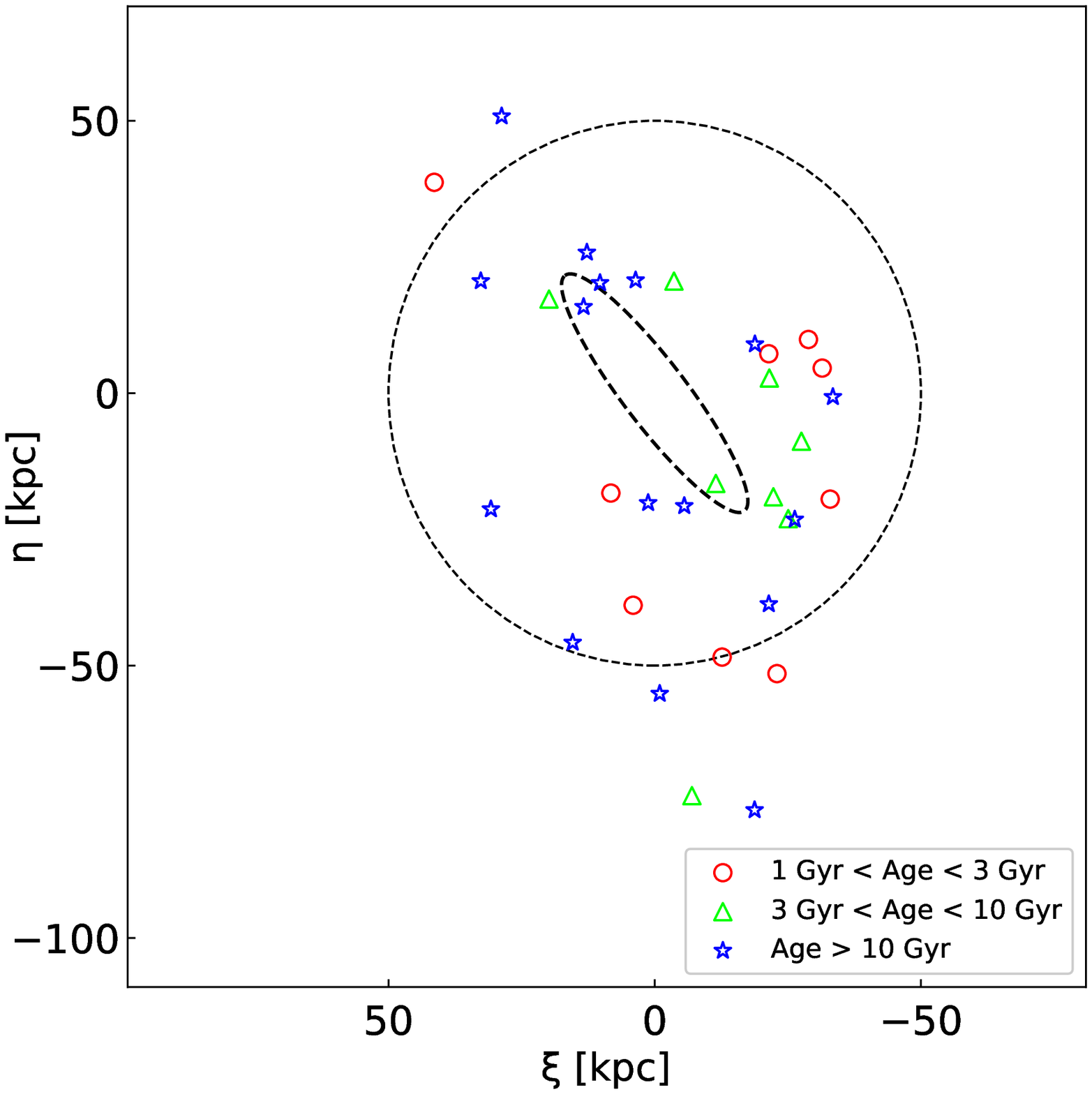}
\includegraphics[width=0.47\textwidth]{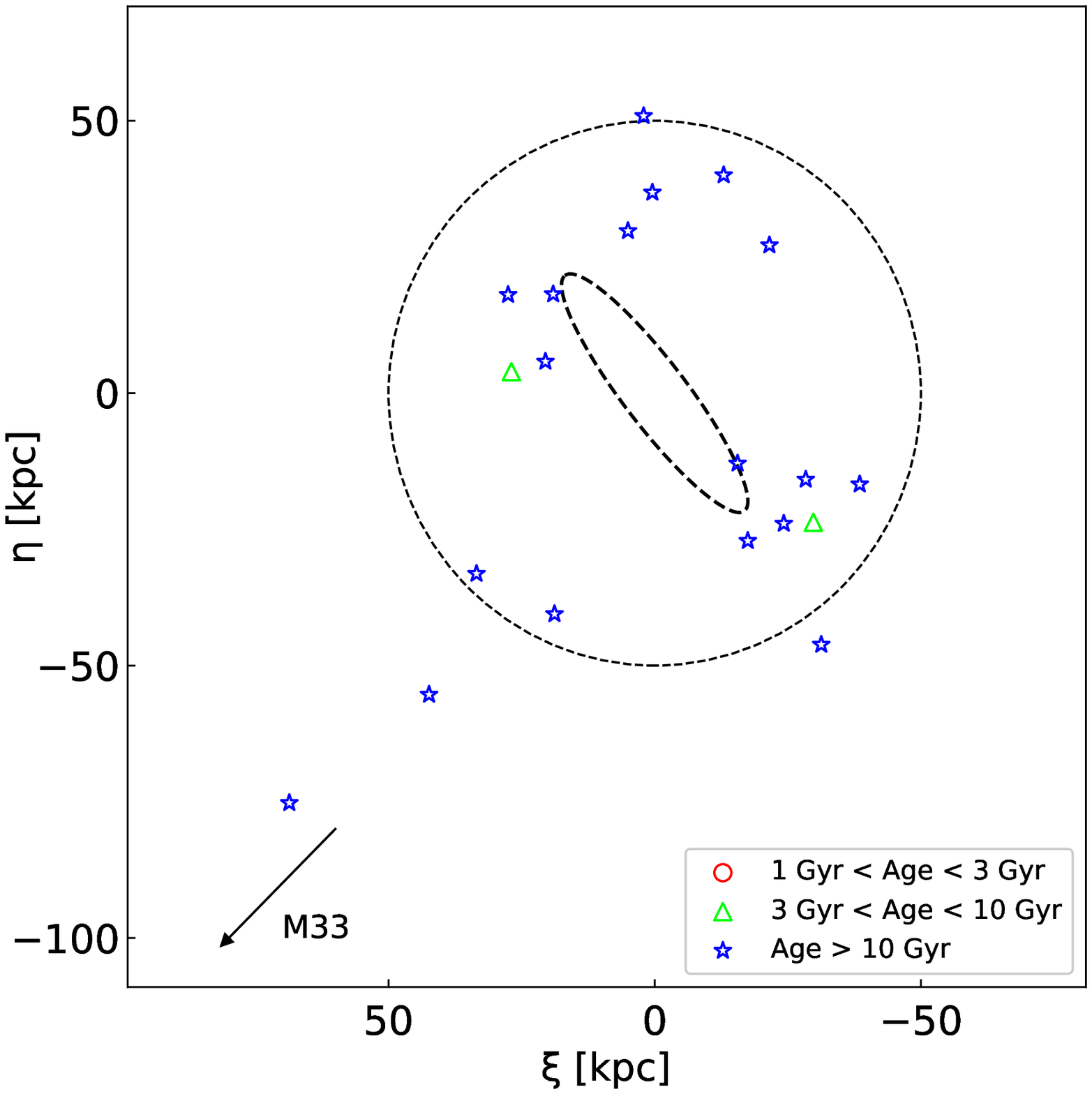}
\caption{Top panel: spatial distribution of GCs with [Fe/H] $>$ $-$1.5. The dashed ellipse has a semimajor axis of 2\o (27 kpc) representing a disk, while the dashed circle lies at a radius of 50 kpc.
Bottom panel: spatial distribution of GCs with [Fe/H] $<$ $-$1.5.}
\label{group.fig}
\end{figure}

We divided our sample GCs into different groups using [Fe/H] and age, and studied the spatial distribution of each group (Figure \ref{group.fig}).
There are some clear features:
(1) All GCs younger than 3 Gyr have metallicities [Fe/H] $>$ $-$1.5, and most of them are located at the south-west region;
(2) Most GCs (18 out of 21) with [Fe/H] $<$ $-$1.5 have an age of 14 Gyr.
(3) There is an extended feature along the direction of M31 minor optical axis and towards M33 for GCs with [Fe/H] $<$ $-$1.5. This feature consists of four GCs (i.e., H22, H27, HEC11, and HEC13), spanning up to 60 kpc. It suggests a possible interaction between M31 and M33 at past;
However, the feature is likely caused by our uneven pointed observations (see Figure \ref{spatial.fig}).

\subsection{Association with Substructures}

\begin{figure}[!htbp]
\center
\includegraphics[width=0.47\textwidth]{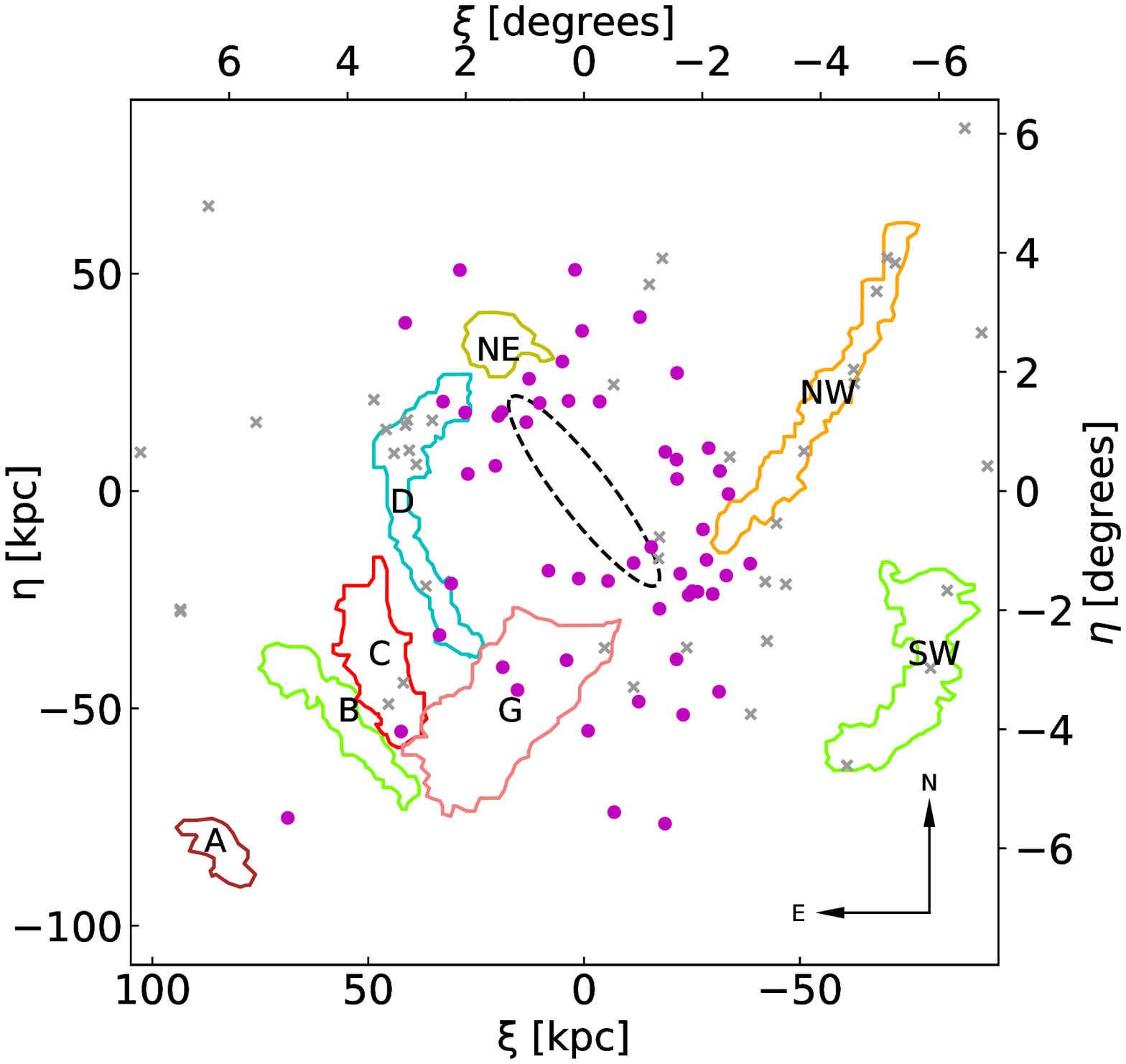}
\caption{The location of the GCs and major substructures in the M31 halo.
The filled circles represent the outer halo GCs with parameter estimates in our study,
while the crosses represent the outer halo GCs without parameter estimates.
The types represent that A: Stream A, B: Stream B, C: Stream C, D: Eastern Arc (Stream D), G: Giant Stream, NE: NE Structure, NW: NW Stream, and SW: SW cloud.
The contours of these substructures are from Figure 3 of \citet{Mackey2010}.
The dashed ellipse has a semimajor axis of 2\o (27 kpc) representing a disk.}
\label{halo.fig}
\end{figure}

The spatial distribution of halo GCs in M31 is anisotropical.
Spatial density maps display a striking association between many GCs and the most luminous substructures in the metal-poor field halo \citep{Mackey2018}.
Kinematic analysis suggests a dynamical
link between the halo GCs and the substructures (i.e., streams or overdensities) that they project onto \citep{Veljanoski2014}.
On the other hand,
many of the surviving dwarf galaxies associated with M31 lie in a thin rotating planar structure \citep{Ibata2013},
and the outer halo GC system appears to be rotating in the same direction as the dwarfs, although with a different rotation axis \citep{Veljanoski2014}.
All of these indicate that a number of outer halo GCs in M31 were accreted from their parent dwarf galaxies.
Recently, \citet{Mackey2018} reported that about 35\%--60\% of remote clusters were accreted into the outskirts of M31 at late times with their parent dwarfs, while more than 40\% of the clusters were accumulated at early times from disrupted primitive satellites. This suggests that the M31 halo system
($R_{\rm p} \geq 25$ kpc) accreted GCs from external galaxies over a Hubble time \citep[see][for details]{Mackey2018}.

To investigate these in more detail, we collected the discovered stellar substructures from previous studies:
the four minor-axis tangent Streams A -- D \citep{Ibata2007};
the Giant Stream \citep{Ibata2001};
the North-East Structure \citep{Zucker2004, Ibata2005};
the North-West Stream \citep{McConnachie2009};
the South-West Cloud \citep{McConnachie2009}.
We collected the contours of these substructures from Figure 3 of \citet{Mackey2010}.
There are totally about 30 GCs (including 8 GCs in our sample) showing spatial coincidence with these halo substructures (Figure \ref{halo.fig}).

We collected the metallicities of substructures from previous studies \citep{Ibata2005,Ibata2007,Ibata2014,Mackey2018}.
The Stream C, Giant Stream, and North-East Structure have a wider range of metallicites than other substructures \citep{Mackey2018}.
The metallicities of GCs show good agreement with those of their specially-associated substructures (Table \ref{associate.tab}).
The metallicity consistency indicates that they have similar stellar populations, which adds further support for the physical association of them.

\begin{table}
\begin{center}
\setlength{\tabcolsep}{0.3em}
\caption{Metallicities of the GCs and their specially-associated substructures.} \vspace{3mm} \label{associate.tab}
\begin{tabular}{lccc}
\hline\noalign{\smallskip}
\multicolumn{2}{c}{Substructures} &    \multicolumn{2}{c}{GC} \\
\cmidrule(l){1-2}\cmidrule(l){3-4}
Name &    [Fe/H] range  &      Name & [Fe/H] \\
    (1)       &    (2)     &    (3)   & (4)  \\
\hline\noalign{\smallskip}
Stream C                          & [$-$2.5, 0]$^a$                           &   HEC13 & $-2.2\pm$0.6 \\
\noalign{\smallskip}
\hline\noalign{\smallskip}
 \multirow{4}{*}{Stream D}        &  \multirow{4}{*}{[$-$2.5,   $-$1.1]$^b$}         &  HEC11 & $-2.2\pm$0.1 \\
                                  &                                      &  H23 & $-1.4\pm$0.1 \\
                                  &                                      &  H24 & $-1.2\pm$0.1 \\
                                  &                                       &  PAndAS-41 & $-1.5\pm$0.1 \\
\hline\noalign{\smallskip}
  \multirow{3}{*}{Giant Stream}   &   \multirow{3}{*}{[$-$2.5, 0]$^a$} &    H19 & $-0.9\pm$0.1 \\
                                 &          &     H22 & $-1.9\pm$0.1 \\
                                 &                                       &   PAndAS-37 &  $-0.4\pm$0.1 \\
\hline
\end{tabular}
\end{center}
{${^a}$ [Fe/H] range is from \citet{Ibata2014}.\\
${^b}$ [Fe/H] range is from \citet{Mackey2018}.}
\end{table}

\subsection{Comparison with M31 Disk GC System}

The GC luminosity function (GCLF) --- the
brightness distribution of GCs in a galaxy,
can be used to constrain the possibilities for GC formation and destruction \citep[see][and references therein]{Nantais2006}.
We determined the GCLFs for M31 halo and disk GCs respectively, using the BATC $g$-band and 2MASS $J$-band magnitudes.
%
We collected 358 disk GCs from \citet{Caldwell2011,Caldwell2016} with $R_{\rm p} \le 20$ kpc,
of which 296 have BATC $g$ magnitudes \citep{Ma2015} and 329 have 2MASS $J$ magnitudes \citep{Wang2014}.
For the halo GCs in our study, 53 and 44 have BATC $g$ and 2MASS $J$ magnitude estimates, respectively.

\begin{figure*}
\center
\includegraphics[width=0.47\textwidth]{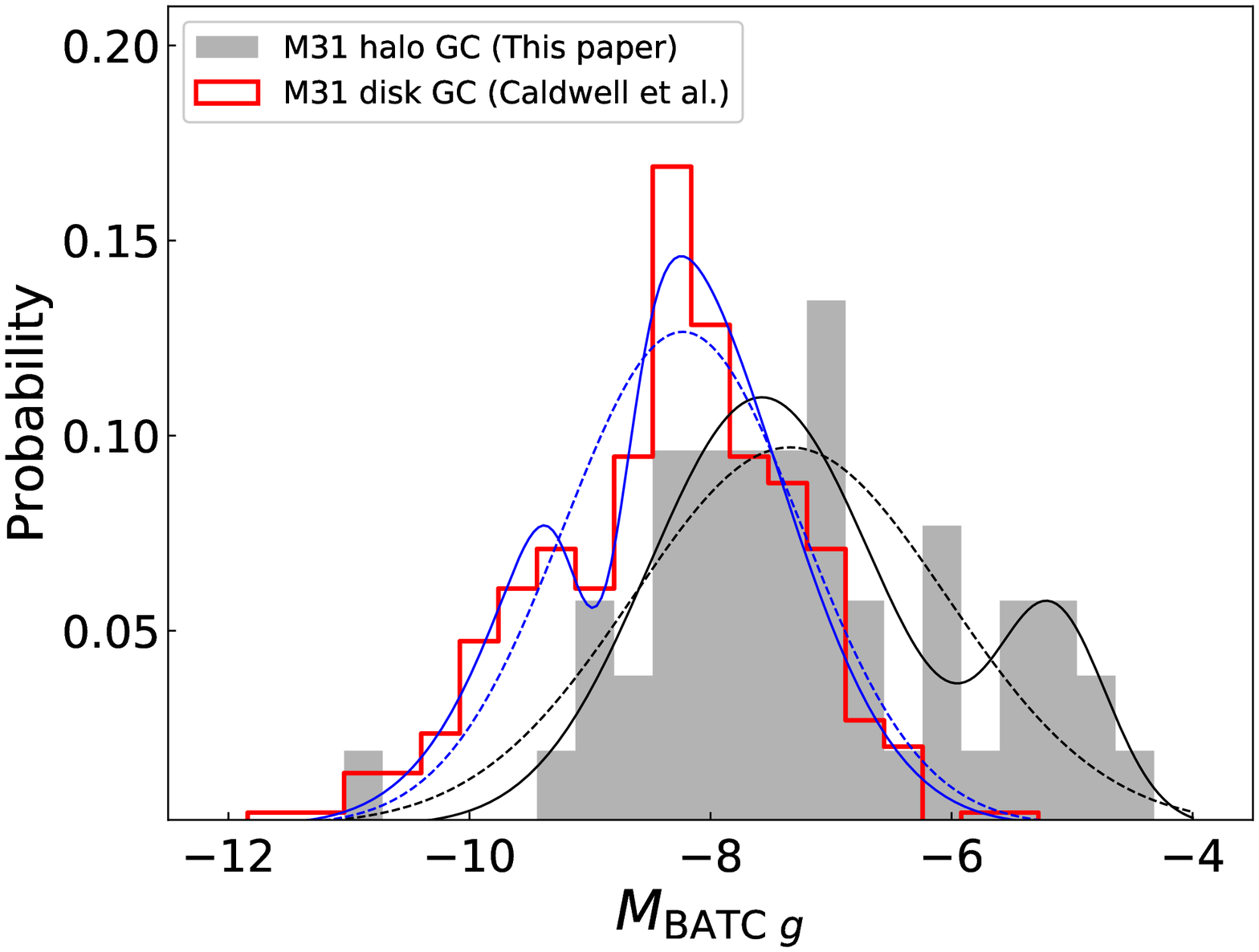}
\includegraphics[width=0.47\textwidth]{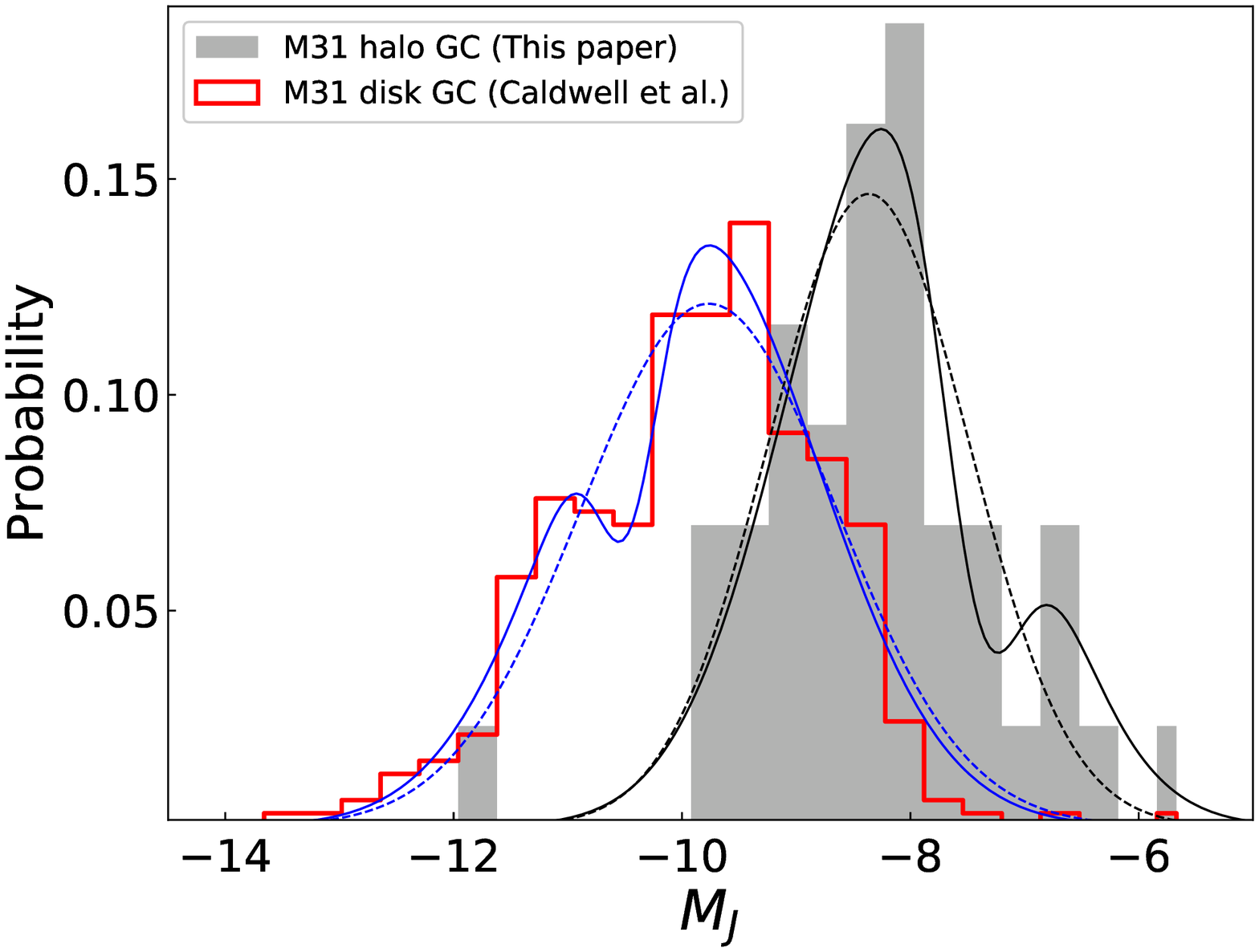}
\caption{Left Panel: luminosity functions of M31 disk and halo GCs, using the BATC $g$ magnitudes. The solid blue and black lines are the double Gaussian fittings to the disk and halo GCLFs, respectively; the dashed blue and black lines are the single Gaussian fittings to the disk and halo GCLFs, respectively.
Right panel: luminosity functions of M31 disk and halo GCs, using the 2MASS $J$ magnitudes.}
\label{comdisklf.fig}
\end{figure*}

We fitted both the GCLFs using one single and double Gaussian functions (Table \ref{diskhalo.tab}), and plotted the results in Figure \ref{comdisklf.fig}.
For the disk GCs, the F-test shows that the double Gaussian fitting is statistically significantly better than the single Gaussian fitting.
The halo GCs show significant bimodality with the BATC $g$ magnitudes, with the probability of the F-test less than 2\%.
It is striking that there are many more faint halo GCs ($M_g > -6$) than the disk ones, and these faint GCs contribute to the fainter part in the GCLF.
\citet{Mackey2010} argued that perhaps up to $\approx$ 80\% of the  GCs in the M31 outer halo were accreted from their parent dwarf galaxies.
Therefore, this bimodality of the GCLF may reflect different origin or evolution environment of these halo GCs in their hosts \citep[see][for details]{Huxor2014,Mackey2018}.

\begin{figure*}
\center
\includegraphics[width=0.47\textwidth]{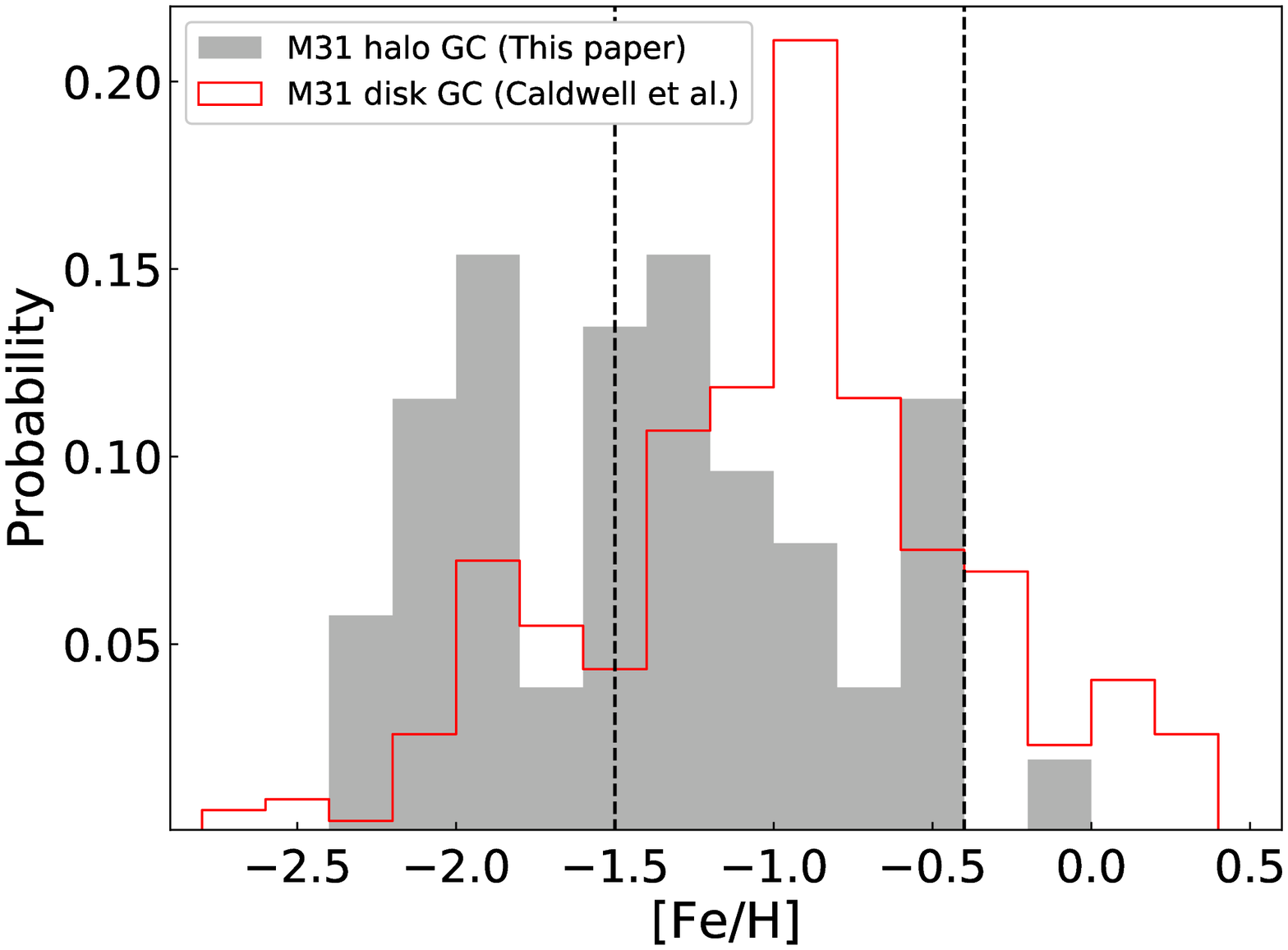}
\includegraphics[width=0.47\textwidth]{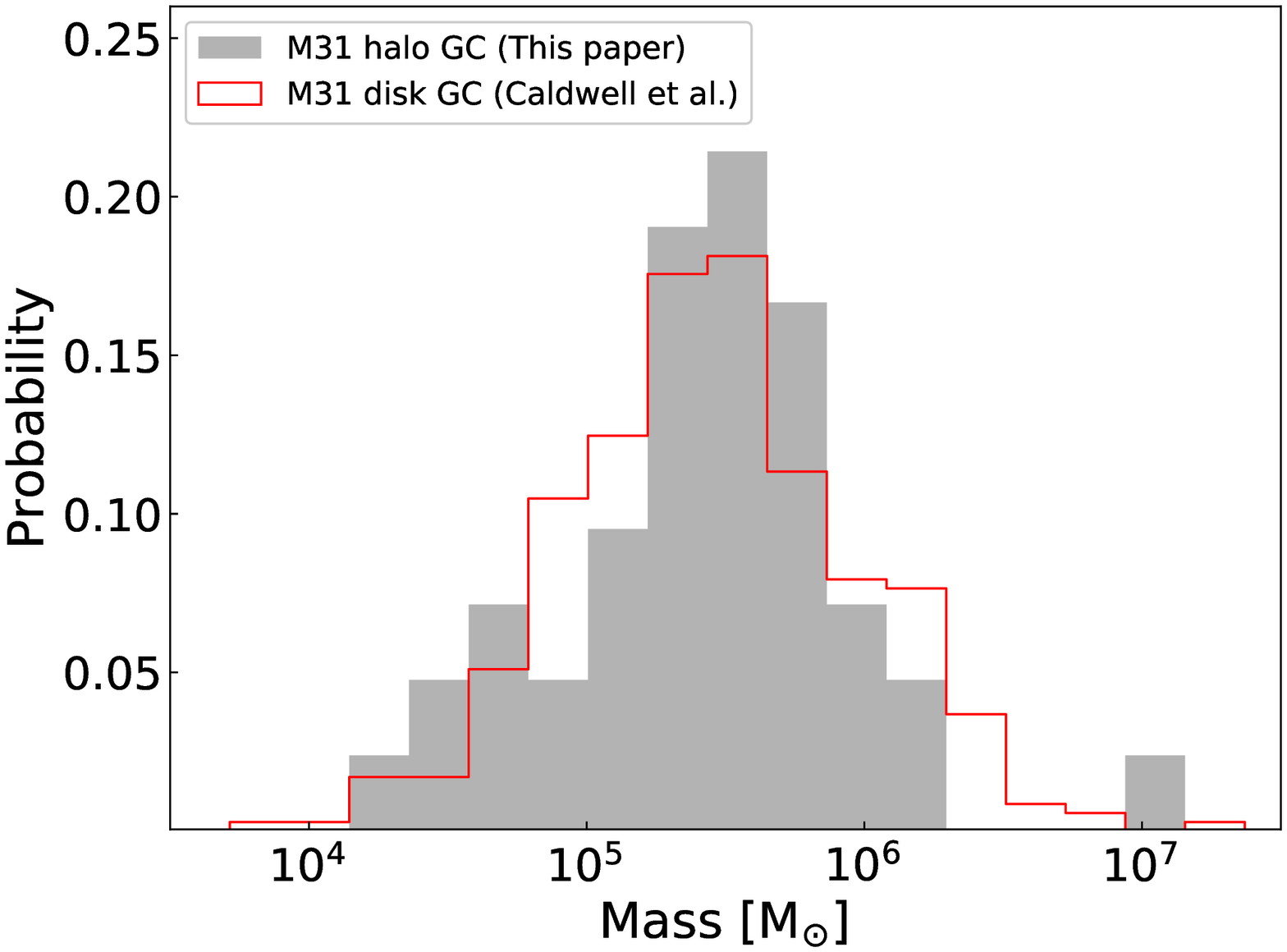}
\caption{Comparison of the metallicities (left panel) and masses (right panel) between the M31 halo and disk GCs.}
\label{comdisk.fig}
\end{figure*}

Previous studies show a bimodal distribution of metallicity of M31 GCs \citep[e.g.,][]{Perrett2002, Puzia2005, Fan2008}.
However, \citet{Caldwell2011,Caldwell2016} argued that the bimodal metallicity distribution is due to the contamination of young disk clusters,
and the M31 distribution can be divided into three major metallicity groups based on their radial distributions.
Figure \ref{comdisk.fig} showed the distributions of metallicity and mass for the outer halo GCs and the disk GCs, respectively. Compared to the disk GCs, there are only two metallicity groups for the outer halo GCs --- one group with intermediate metallicity ($-1.5 \geq$ [Fe/H] $< -0.4$) and one metal-poor group ([Fe/H] $<-1.5$).
The masses of disk GCs span a wider range than halo GCs, both in the lower and upper end.
The cluster G001 is the most massive one among the halo GCs, and it is also one of the most massive GCs in M31.
It is likely the remnant core of a dwarf galaxy which lost most of its envelope through tidal interactions with M31 \citep{Meylan1997,Meylan2001,ma2007a}.

\begin{table*}
\begin{center}
\caption{Fitting results of the luminosity functions of M31 disk and halo GC systems.} \vspace{3mm} \label{diskhalo.tab}
\begin{tabular}{lccccccc}
\hline\noalign{\smallskip}
 \multirow{2}{*}{GC}  & \multicolumn{4}{c}{Double Gaussian} &  \multicolumn{2}{c}{Single Gaussian}  &   \multirow{2}{*}{P$^{a}$} \\
  \cmidrule(l){2-5}\cmidrule(l){6-7}
 &    $\mu_1$  & $\sigma_1$ & $\mu_2$  &  $\sigma_2$ &  $\mu$  & $\sigma$  & \\
   (1) & (2) &(3) &(4) &(5) &   (6) & (7)   & (8)\\
   \hline\noalign{\smallskip}
    & \multicolumn{6}{c}{BATC $g$} \\
\hline\noalign{\smallskip}
Disk   & $ -8.41\pm0.05 $ & $ 0.96\pm0.04 $ & $-8.91\pm0.04 $ & $ -0.26\pm0.05 $ &  $ -8.23\pm0.07 $ & $ 0.99\pm0.07 $  &  0.01\%  \\
Halo  & $-7.57\pm0.11$ & $0.94\pm0.12$ & $-5.17\pm0.16$ & $0.45\pm0.16$ &     $-7.34\pm0.16$ & $1.3\pm0.16$  &   1.72\% \\
\hline\noalign{\smallskip}
& \multicolumn{6}{c}{2MASS $J$} \\
\hline\noalign{\smallskip}
Disk   &    $-9.9\pm0.06 $ & $1.1\pm0.05 $ & $-10.46\pm0.06 $ & $0.27\pm0.06 $ &      $ -9.77\pm0.07 $ & $ 1.12\pm0.07 $  & 0.17\% \\
Halo  &    $-8.09\pm0.16$ & $0.97\pm0.09$ & $-7.4\pm0.11$  &  $0.35\pm0.15$ &      $-8.36\pm0.09$ & $0.88\pm0.09$  &  9.24\%   \\
\hline
\end{tabular}
\end{center}
{${^a}$ P is the probability from the F-test. If the P value is low (e.g., 5\%),
we can conclude that the double Gaussian model is statistically significantly better than the single Gaussian model.}
\end{table*}

\subsection{Comparison with Galactic GC System}

It has been known that there are differences between the MW and M31 GC systems:
(1) M31 contains more than twice the total number of GCs in the MW \citep{Galleti2004,Peacock2010,Caldwell2016};
%
(2) The metallicity distribution of the MW GCs is bimodal, with peaks at [Fe/H] $\approx$ $-$1.5 and $-$0.6 dex \citep{Zinn1985},
while that of M31 GCs includes three components, covering [Fe/H] from $-$3 to 0.5 \citep{Elson1988,Caldwell2016};
(3) M31 has a subcomponent of GCs (mostly metal-rich) that follow closely the thin-disk kinematics (e.g., rotation, velocity dispersion), and such a GC system has not yet been found in the MW \citep{Morrison2004,Caldwell2016}.

However, there are also remarkable similarities in the chemodynamical properties between the M31 and MW GC systems. Both of the GC systems are dominated by the metal-poor group, which is spherically distributed and has a high velocity dispersion \citep{Caldwell2016}.
In addition, there is a planar subgroup of dwarf galaxies in the M31 \citep{Ibata2013}, and they appear to be rotating in the same direction as the outer halo GCs \citep{Veljanoski2014}.
Similar plane has also been found in the MW \citep{Metz2007, Keller2012,Pawlowski2012}.

\begin{figure*}
\center
\includegraphics[width=0.47\textwidth]{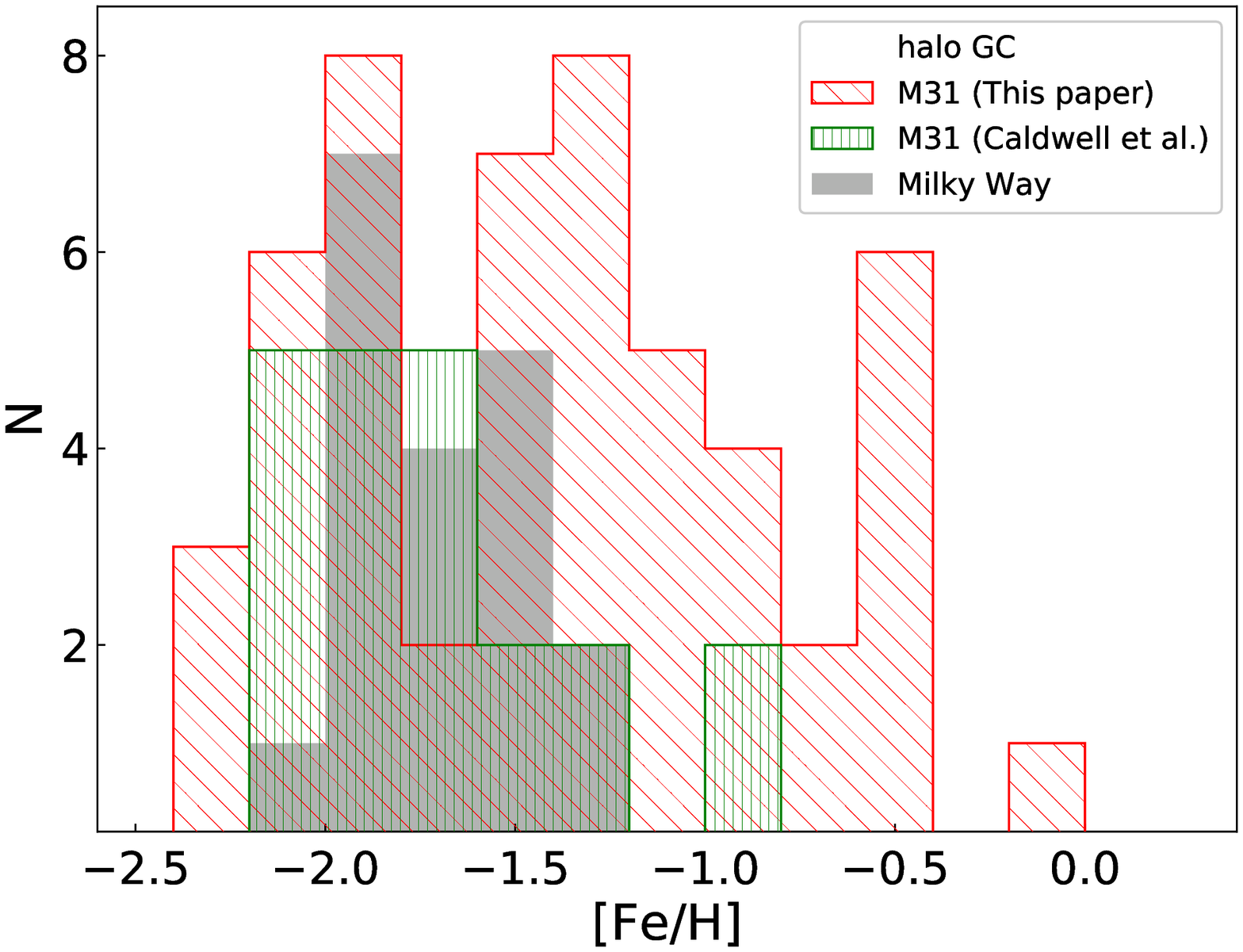}
\includegraphics[width=0.47\textwidth]{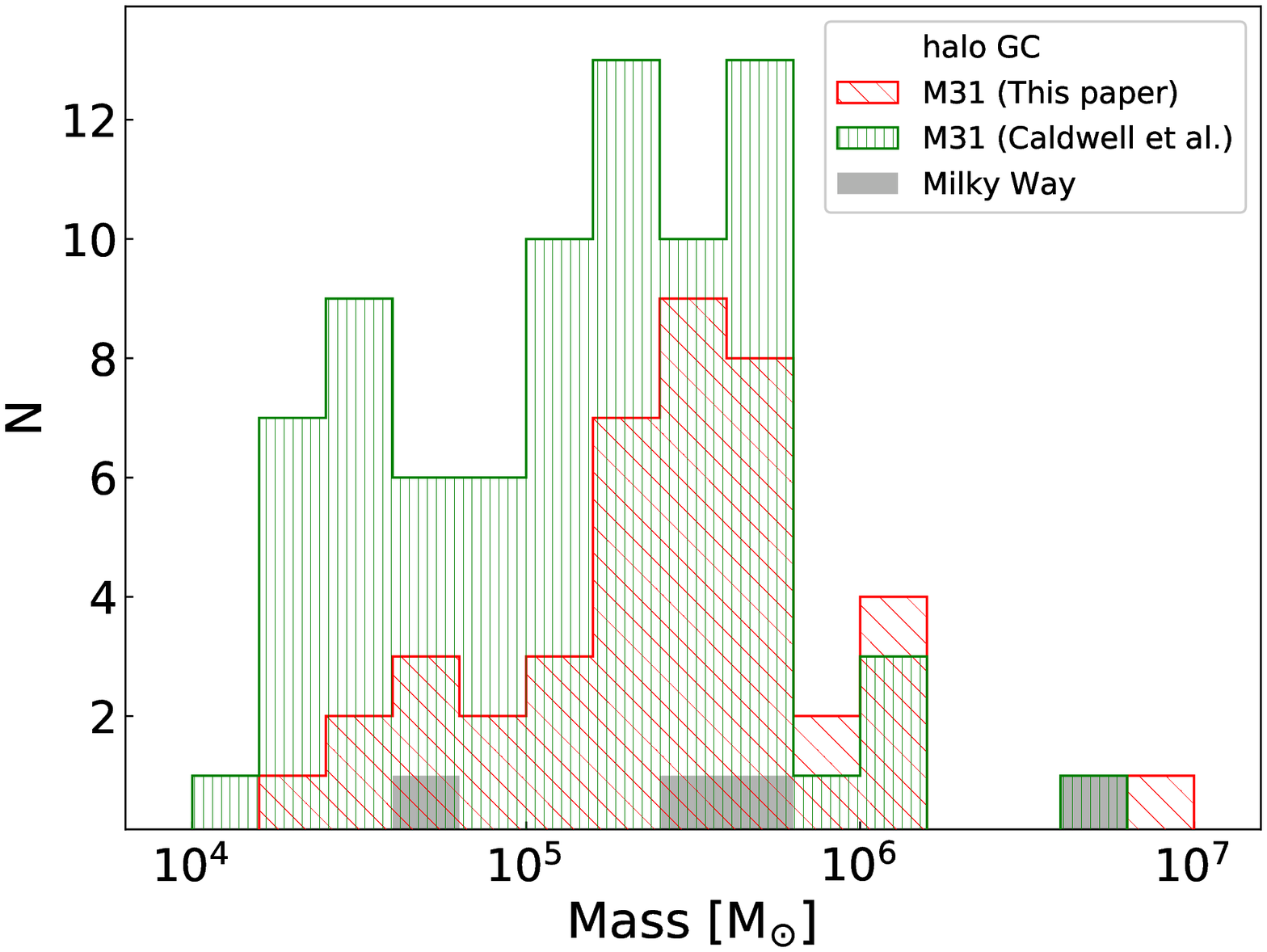}
\caption{Comparison of the metallicities (left panel) and masses (right panel) between the M31 and Galactic halo GCs.}
\label{commw.fig}
\end{figure*}

By defining the outer halo with $R_{\rm p} >$ 20 kpc, it is intriguing to see that
M31 has more than 100 outer halo GCs ($N =$ 113), which is $\approx$ 6 times the number of halo GCs in the MW ($N =$ 19).
Considering that the total number of M31 GCs is approximately three times more numerous than that of the MW, it seems that M31 contains too many halo GCs.
Figure \ref{commw.fig} showed the comparison of metallicity and mass between M31 and Galactic halo GCs. For the Galactic outer halo GCs, we obtained the metallicities from \citet{Harris1996} and masses from \citet{Pryor1993} and \citet{Kimmig2015}.
The metallicity of the GCs in M31 halo spans a wider range than that of GCs in the MW halo;
the M31 halo GCs from \citet{Caldwell2016} present a similar metallicity range with the Galactic halo GCs, but their number is quite small ($\approx$ 20).
For the MW halo, previous studies have shown that it is very metal-poor \citep{Ryan1991,Chiba2000}, with a median metallicity of [Fe/H] $=$ $-$1.6.
In contrast, M31 halo includes a group with intermediate metallicity ([Fe/H] $=$ $-$0.6) and a metal-rich component with [Fe/H] $=$ $-$0.2 \citep{Mould1986,Bellazzini2003}.
The minor axis halo fields of M31 show a de Vaucouleurs $R^{1/4}$-law density profile \citep{Durrell2004} and intermediate metallicity ([Fe/H] $\approx$ $-$0.5), different with the $R^{-3.5}$ shape of the Galactic metal-poor halo \citep{Chiba2000, Ibata2005}.
However, using a group of kinematically selected stars,
\citet{Chapman2006} showed that the M31 halo population has [Fe/H] $\approx$ $-$1.4 with a dispersion of 0.2 dex, very similar to that found in the MW; this indicates the metal-rich component does not have halo-like kinematics \citep{Ibata2007}.
These GCs with [Fe/H] > $-$1 are consistent with the component with intermediate metallicities discovered before.
The host systems of these GCs may have undergone efficient star formation and chemical enrichment, indicating that they were accreted into the M31 halo at late times \citep{Mackey2018}.
All of these are suggestive of an active merger history of M31 during the halo formation \citep{Ibata2005}.

\section{SUMMARY}

Previous studies showed that the MW has undergone a low amount of merging, whereas M31 appears to be much more different \citep[e.g.,][]{Hammer2007}.
This is consistent with the unusual features  (e.g., young stellar population, rotational signature of substructures)
observed in the M31 outer regions \citep{Ferguson2002,Ibata2005,Richardson2008}, which
are thought as evidences of a history with recent merging \citep{McConnachie2009}.
Similarly, the M31 outer halo GC system shows a strong rotation around the minor optical axis of M31 \citep{Veljanoski2014},
while the Galactic halo GC population exhibits only slight net rotation \citep[e.g.,][]{Brodie2006, Deason2011}.
Stellar population studies of halo GCs may provide more straight evidence of the merging/accretion events.

In this paper, we determined the metallicities, ages, and masses for 53 GCs in the M31 outer halo,
by comparing the multicolour photometry with theoretical SSP models.
The multicolour photometric data are obtained from the {\sl GALEX} FUV and NUV, SDSS $ugriz$,
15 intermediate-band filters of BATC, and 2MASS $JHK_{\rm s}$, which constitute the SEDs covering 1538 -- 20,000 \AA.
In general, our results based on SED fitting are consistent with previous studies.

We studied the spatial distribution of the halo GCs, and found
no clear trend of metallicity and mass with the de-projected radius.
We also found that the halo GCs with age younger than $\approx$ 8 Gyr are mostly located at the de-projected radii around 100 kpc,
but this may be due to a selection effect;
%
and that most metal-poor GCs ([Fe/H] $<$ $-$1.5) have an age of 14 Gyr.
Our results showed that there is an extended feature towards M33 for these metal-poor GCs, which is along the direction of M31 minor optical axis, consisting of four GCs: H22, H27, HEC11, and HEC13.
However, the feature is likely a consequence of our uneven pointed observations.
In our sample, there are eight GCs showing spatial coincidence with these halo substructures,
and these GCs have consistent metallicities with their spatially-associated substructures.
Together with the dynamical link between the halo GCs and the substructures found in previous kinematic analysis,
our results provide further evidence of the physical association of them.

Our results showed that both the disk and halo GCs in M31 show a bimodal luminosity distribution,
and that there are many more faint halo GCs than the disk ones, the former of which contribute to the fainter part in the halo GCLF.
Because most of the GCs in M31 outer halo were accreted from dwarf galaxies, this bimodality of the GCLF may reflect different origin or evolution environment of these halo GCs.
The M31 disk GCs includes three metallicity components, covering from $-$3 to 0.5,
while there are only two groups for the outer halo GCs --- one group with intermediate metallicity ($-1.5 \geq$ [Fe/H] $< -0.4$) and one metal-poor group ([Fe/H] $<-1.5$).

The total number of M31 GCs is approximately three times more numerous than that of the MW, however,
M31 has more than 100 outer halo GCs ($N =$ 113), which is $\approx 6$ times the number of halo GCs in the MW ($N =$ 19).
It seems that M31 contains too many halo GCs.
%
The metallicity of the GCs in M31 halo spans a wider range than that of GCs in the MW halo, since the Galactic halo GCs are mostly metal-poor.
Both the numerous halo GCs and the higher-metallicity component are suggestive of an active merger history of M31 during the halo formation.

\begin{acknowledgements}
We especially thank the anonymous referee for his/her thorough report and helpful comments
and suggestions that have significantly improved the paper.
We thank Dr. Zhou Fan for helpful discussions.
This work was supported by the Chinese National Natural Science Foundation grant Nos. 11603035, 11873053, 11433005, 11273028, 11333004, 2016YFA0400800, and 11425313.
\end{acknowledgements}





\begin{thebibliography}{}

\bibitem[Agar
\& Barmby(2013)]{Agar2013} Agar, J.~R.~R., \& Barmby, P.\ 2013, \aj, 146, 135


\bibitem[Alves-Brito et al.(2009)]{Alves-Brito2009} Alves-Brito, A., Forbes, D.~A., Mendel, J.~T., Hau, G.~K.~T., \& Murphy, M.~T.\ 2009, \mnras, 395, L34



\bibitem[Anders et al.(2004)]{Anders2004} Anders, P., Bissantz, N.,
Fritze-v.~Alvensleben, U., \& de Grijs, R.\ 2004, \mnras, 347, 196


\bibitem[Barmby et al.(2000)]{Barmby2000} Barmby, P., Huchra, J.~P.,
Brodie, J.~P., et al.\ 2000, \aj, 119, 727


\bibitem[Barmby et al.(2007)]{Barmby2007} Barmby, P., McLaughlin, D.~E.,
Harris, W.~E., Harris, G.~L.~H., \& Forbes, D.~A.\ 2007, \aj, 133, 2764





\bibitem[Bellazzini et al.(2003)]{Bellazzini2003} Bellazzini, M., Cacciari,
C., Federici, L., Fusi Pecci, F., \& Rich, M.\ 2003, \aap, 405, 867

\bibitem[Bertelli et al.(1994)]{Bertelli1994} Bertelli, G., Bressan, A.,
Chiosi, C., Fagotto, F., \& Nasi, E.\ 1994, \aaps, 106, 275

\bibitem[Bertin
\& Arnouts(1996)]{Bertin1996} Bertin, E., \& Arnouts, S.\ 1996, \aaps, 117, 393

\bibitem[Brodie
\& Strader(2006)]{Brodie2006} Brodie, J.~P., \& Strader, J.\ 2006, \araa, 44, 193


\bibitem[Bruzual
\& Charlot(2003)]{Bruzual2003} Bruzual, G., \& Charlot, S.\ 2003, \mnras, 344, 1000




\bibitem[Caldwell et al.(2009)]{Caldwell2009} Caldwell, N., Harding, P.,
Morrison, H., et al.\ 2009, \aj, 137, 94

\bibitem[Caldwell et al.(2016)]{Caldwell2016} Caldwell, N., \& Romanowsky, A.~J.\ 2016, \apj, 824, 42

\bibitem[Caldwell et al.(2011)]{Caldwell2011} Caldwell, N., Schiavon, R.,
Morrison, H., Rose, J.~A., \& Harding, P.\ 2011, \aj, 141, 61


\bibitem[Cardelli et al.(1989)]{Cardelli1989} Cardelli, J.~A., Clayton,
G.~C., \& Mathis, J.~S.\ 1989, \apj, 345, 245

\bibitem[Chapman et al.(2006)]{Chapman2006} Chapman, S.~C., Ibata, R.,
Lewis, G.~F., et al.\ 2006, \apj, 653, 255



\bibitem[Chen et al.(2016)]{Chen2016} Chen, B., Liu, X., Xiang, M., et al.\
2016, \aj, 152, 45

\bibitem[Chiba
\& Beers(2000)]{Chiba2000} Chiba, M., \& Beers, T.~C.\ 2000, \aj, 119, 2843

\bibitem[Colucci et al.(2014)]{Colucci2014} Colucci, J.~E., Bernstein,
R.~A., \& Cohen, J.~G.\ 2014, \apj, 797, 116

\bibitem[C{\^o}t{\'e} et al.(2000)]{Cote2000} C{\^o}t{\'e}, P., Marzke,
R.~O., West, M.~J., \& Minniti, D.\ 2000, \apj, 533, 869

\bibitem[Da Costa
\& Armandroff(1995)]{DaCosta1995} Da Costa, G.~S., \& Armandroff, T.~E.\ 1995, \aj, 109, 2533

\bibitem[Deason et al.(2011)]{Deason2011} Deason, A.~J., Belokurov, V.,
\& Evans, N.~W.\ 2011, \mnras, 411, 1480

\bibitem[de Grijs et al.(2003)]{deGrijs2003} de Grijs, R.,
Fritze-v.~Alvensleben, U., Anders, P., et al.\ 2003, \mnras, 342, 259


\bibitem[de Jong(1996)]{deJong1996} de Jong, R.~S.\ 1996, \aap, 313, 377


\bibitem[di Tullio Zinn
\& Zinn(2014)]{dTZZ2014} di Tullio Zinn, G., \& Zinn, R.\ 2014, \aj, 147, 90


\bibitem[di Tullio Zinn
\& Zinn(2013)]{dTZZ2013} di Tullio Zinn, G., \& Zinn, R.\ 2013, \aj, 145, 50

\bibitem[Durrell et al.(2004)]{Durrell2004} Durrell, P.~R., Harris, W.~E.,
\& Pritchet, C.~J.\ 2004, \aj, 128, 260

\bibitem[Elson
\& Walterbos(1988)]{Elson1988} Elson, R.~A., \& Walterbos, R.~A.~M.\ 1988, \apj, 333, 594


\bibitem[Fan et al.(1996)]{Fan1996} Fan, X., Burstein, D., Chen, J.-S., et
al.\ 1996, \aj, 112, 628

\bibitem[Fan et al.(2010a)]{Fan2010a} Fan, Z., de Grijs, R.,
\& Zhou, X.\ 2010a, \apj, 725, 200

\bibitem[Fan et al.(2008)]{Fan2008} Fan, Z., Ma, J., de Grijs, R.,
\& Zhou, X.\ 2008, \mnras, 385, 1973

\bibitem[Fan et al.(2009)]{Fan2009} Fan, Z., Ma, J.,
\& Zhou, X.\ 2009, Research in Astronomy and Astrophysics, 9, 993


\bibitem[Fan et al.(2010b)]{Fan2010b} Fan, Z., Ma, J., Zhou, X.,
\& Jiang, Z.\ 2010b, \pasp, 122, 636


\bibitem[Fan
\& Wang(2017)]{Fan2017} Fan, Z., \& Wang, S.\ 2017, \apss, 362, 193

\bibitem[Ferguson et al.(2002)]{Ferguson2002} Ferguson, A.~M.~N., Irwin,
M.~J., Ibata, R.~A., Lewis, G.~F., \& Tanvir, N.~R.\ 2002, \aj, 124, 1452

\bibitem[Galleti et al.(2004)]{Galleti2004} Galleti, S., Federici, L.,
Bellazzini, M., Fusi Pecci, F., \& Macrina, S.\ 2004, \aap, 416, 917






\bibitem[Hammer et al.(2007)]{Hammer2007} Hammer, F., Puech, M., Chemin,
L., Flores, H., \& Lehnert, M.~D.\ 2007, \apj, 662, 322


\bibitem[Harris(1996)]{Harris1996} Harris, W.~E.\ 1996, \aj, 112, 1487


\bibitem[Harris et al.(2017)]{Harris2017} Harris, W.~E., Ciccone, S.~M.,
Eadie, G.~M., et al.\ 2017, \apj, 835, 101




\bibitem[Huchra et al.(1991)]{Huchra1991} Huchra, J.~P., Brodie, J.~P.,
\& Kent, S.~M.\ 1991, \apj, 370, 495


\bibitem[Huxor et al.(2011)]{Huxor2011} Huxor, A.~P., Ferguson, A.~M.~N.,
Tanvir, N.~R., et al.\ 2011, \mnras, 414, 770


\bibitem[Huxor et al.(2014)]{Huxor2014} Huxor, A.~P., Mackey, A.~D.,
Ferguson, A.~M.~N., et al.\ 2014, \mnras, 442, 2165


\bibitem[Huxor et al.(2008)]{Huxor2008} Huxor, A.~P., Tanvir, N.~R.,
Ferguson, A.~M.~N., et al.\ 2008, \mnras, 385, 1989




\bibitem[Ibata et al.(2005)]{Ibata2005} Ibata, R., Chapman, S., Ferguson,
A.~M.~N., et al.\ 2005, \apj, 634, 287


\bibitem[Ibata et al.(1994)]{Ibata1994} Ibata, R.~A., Gilmore, G.,
\& Irwin, M.~J.\ 1994, \nat, 370, 194


\bibitem[Ibata et al.(2007)]{Ibata2007} Ibata, R., Martin, N.~F., Irwin,
M., et al.\ 2007, \apj, 671, 1591


\bibitem[Ibata et al.(2001)]{Ibata2001} Ibata, R., Irwin, M., Lewis, G.,
Ferguson, A.~M.~N., \& Tanvir, N.\ 2001, \nat, 412, 49


\bibitem[Ibata et al.(2013)]{Ibata2013} Ibata, R.~A., Lewis, G.~F., Conn,
A.~R., et al.\ 2013, \nat, 493, 62


\bibitem[Ibata et al.(2014)]{Ibata2014} Ibata, R.~A., Lewis, G.~F.,
McConnachie, A.~W., et al.\ 2014, \apj, 780, 128


\bibitem[Kang et al.(2012)]{Kang2012} Kang, Y., Rey, S.-C., Bianchi, L., et
al.\ 2012, \apjs, 199, 37


\bibitem[Kaviraj et al.(2007)]{Kaviraj2007} Kaviraj, S., Rey, S.-C., Rich,
R.~M., Yoon, S.-J., \& Yi, S.~K.\ 2007, \mnras, 381, L74

\bibitem[Keller et al.(2012)]{Keller2012} Keller, S.~C., Mackey, D.,
\& Da Costa, G.~S.\ 2012, \apj, 744, 57

\bibitem[Kent(1989)]{ken89}Kent, S. 1989, AJ, 97, 1614


\bibitem[Kimmig et al.(2015)]{Kimmig2015} Kimmig, B., Seth, A., Ivans,
I.~I., et al.\ 2015, \aj, 149, 53




\bibitem[Ma et al.(2009)]{Ma2009} Ma, J., Fan, Z., de Grijs, R., et al.\
2009, \aj, 137, 4884

\bibitem[Ma et al.(2012)]{Ma2012} Ma, J., Wang, S., Wu, Z., et al.\ 2012,
\aj, 143, 29

\bibitem[Ma et al.(2015)]{Ma2015} Ma, J., Wang, S., Wu, Z., et al.\ 2015,
\aj, 149, 56

\bibitem[Ma et al.(2007a)]{ma2007a}Ma, J.; de Grijs, R.; Chen, D, et al.\
2007a, \mnras, 376, 1621

\bibitem[Ma et al.(2007b)]{Ma2007b} Ma, J., Yang, Y., Burstein, D., et al.\
2007b, \apj, 659, 359

\bibitem[Mackey et al.(2018)]{Mackey2018} Mackey, D., Ferguson, A., Huxor,
A., et al.\ 2018, arXiv:1810.10719

\bibitem[Mackey et al.(2007)]{Mackey2007} Mackey, A.~D., Huxor, A.,
Ferguson, A.~M.~N., et al.\ 2007, \apjl, 655, L85


\bibitem[Mackey et al.(2006)]{Mackey2006} Mackey, A.~D., Huxor, A.,
Ferguson, A.~M.~N., et al.\ 2006, \apjl, 653, L105


\bibitem[Mackey et al.(2010)]{Mackey2010} Mackey, A.~D., Huxor, A.~P.,
Ferguson, A.~M.~N., et al.\ 2010, \apjl, 717, L11




\bibitem[Martin et al.(2004)]{Martin2004} Martin, N.~F., Ibata, R.~A.,
Bellazzini, M., et al.\ 2004, \mnras, 348, 12

\bibitem[Mateu et al.(2009)]{Mateu2009} Mateu, C., Vivas, A.~K., Zinn, R.,
Miller, L.~R., \& Abad, C.\ 2009, \aj, 137, 4412



\bibitem[McConnachie et al.(2018)]{McConnachie2018} McConnachie, A.~W.,
Ibata, R., Martin, N., et al.\ 2018, arXiv:1810.08234

\bibitem[McConnachie et al.(2005)]{McConnachie2005} McConnachie, A.~W.,
Irwin, M.~J., Ferguson, A.~M.~N., et al.\ 2005, \mnras, 356, 979


\bibitem[McConnachie et al.(2009)]{McConnachie2009} McConnachie, A.~W.,
Irwin, M.~J., Ibata, R.~A., et al.\ 2009, \nat, 461, 66

\bibitem[Metz et al.(2007)]{Metz2007} Metz, M., Kroupa, P.,
\& Jerjen, H.\ 2007, \mnras, 374, 1125


\bibitem[Meylan et al.(2001)]{Meylan2001} Meylan, G., Sarajedini, A.,
Jablonka, P., et al.\ 2001, \aj, 122, 830


\bibitem[Meylan
\& Heggie(1997)]{Meylan1997} Meylan, G., \& Heggie, D.~C.\ 1997, \aapr, 8, 1

\bibitem[Morrison et al.(2004)]{Morrison2004} Morrison, H.~L., Harding, P.,
Perrett, K., \& Hurley-Keller, D.\ 2004, \apj, 603, 87


\bibitem[Mould
\& Kristian(1986)]{Mould1986} Mould, J., \& Kristian, J.\ 1986, \apj, 305, 591

\bibitem[Nantais et al.(2006)]{Nantais2006} Nantais, J.~B., Huchra, J.~P.,
Barmby, P., Olsen, K.~A.~G., \& Jarrett, T.~H.\ 2006, \aj, 131, 1416

\bibitem[Pawlowski et al.(2012)]{Pawlowski2012} Pawlowski, M.~S.,
Pflamm-Altenburg, J., \& Kroupa, P.\ 2012, \mnras, 423, 1109


\bibitem[Peacock et al.(2010)]{Peacock2010} Peacock, M.~B., Maccarone,
T.~J., Knigge, C., et al.\ 2010, \mnras, 402, 803


\bibitem[Peacock et al.(2011)]{Peacock2011} Peacock, M.~B., Zepf, S.~E.,
Maccarone, T.~J., \& Kundu, A.\ 2011, \apj, 737, 5


\bibitem[Peng et al.(2006)]{Peng2006} Peng, E.~W., Jord{\'a}n, A.,
C{\^o}t{\'e}, P., et al.\ 2006, \apj, 639, 95



\bibitem[Perrett et al.(2002)]{Perrett2002} Perrett, K.~M., Bridges, T.~J.,
Hanes, D.~A., et al.\ 2002, \aj, 123, 2490

\bibitem[Pryor
\& Meylan(1993)]{Pryor1993} Pryor, C., \& Meylan, G.\ 1993, Structure and Dynamics of Globular Clusters, 50, 357


\bibitem[Puzia et al.(2005)]{Puzia2005} Puzia, T.~H., Perrett, K.~M.,
\& Bridges, T.~J.\ 2005, \aap, 434, 909



\bibitem[Rich et al.(2005)]{Rich2005} Rich, R.~M., Corsi, C.~E., Cacciari, C., et al.\ 2005, \aj, 129, 2670





\bibitem[Richardson et al.(2008)]{Richardson2008} Richardson, J.~C.,
Ferguson, A.~M.~N., Johnson, R.~A., et al.\ 2008, \aj, 135, 1998



\bibitem[Ryan
\& Norris(1991)]{Ryan1991} Ryan, S.~G., \& Norris, J.~E.\ 1991, \aj, 101, 1865

\bibitem[Sakari et al.(2015)]{Sakari2015} Sakari, C.~M., Venn, K.~A., Mackey, D., et al.\ 2015, \mnras, 448, 1314


\bibitem[Salpeter(1955)]{Salpeter1955} Salpeter, E.~E.\ 1955, \apj, 121,
161


\bibitem[Schlafly
\& Finkbeiner(2011)]{Schlafly2011} Schlafly, E.~F., \& Finkbeiner, D.~P.\ 2011, \apj, 737, 103


\bibitem[Stetson(1987)]{Stetson1987} Stetson, P.~B.\ 1987, \pasp, 99, 191

\bibitem[Veljanoski et al.(2014)]{Veljanoski2014} Veljanoski, J., Mackey,
A.~D., Ferguson, A.~M.~N., et al.\ 2014, \mnras, 442, 2929

\bibitem[Wang et al.(2010)]{Wang2010} Wang, S., Fan, Z., Ma, J., de Grijs,
R., \& Zhou, X.\ 2010, \aj, 139, 1438



\bibitem[Wang et al.(2014)]{Wang2014} Wang, S., Ma, J., Wu, Z.,
\& Zhou, X.\ 2014, \aj, 148, 4

\bibitem[White
\& Rees(1978)]{White1978} White, S.~D.~M., \& Rees, M.~J.\ 1978, \mnras, 183, 341

\bibitem[Zheng et al.(1999)]{Zheng1999} Zheng, Z., Shang, Z., Su, H., et
al.\ 1999, \aj, 117, 2757


\bibitem[Zinn(1985)]{Zinn1985} Zinn, R.\ 1985, \apj, 293, 424


\bibitem[Zucker et al.(2004)]{Zucker2004} Zucker, D.~B., Kniazev, A.~Y.,
Bell, E.~F., et al.\ 2004, \apjl, 612, L117


\end{thebibliography}
\end{document}